\documentclass[twocolumn]{aastex631}

\usepackage{natbib}
\usepackage{amsmath}

\hypersetup{linkcolor=blue,citecolor=blue,filecolor=cyan,urlcolor=magenta}

\providecommand{\dif}{\mathrm{d}} \def\d{\dif}

\def\Msun{\mskip3mu{\rm M}_\odot}

\def\RS{\Sigma}

\newcommand{\beq}{\begin{equation}}
\newcommand{\eeq}{\end{equation}}
\newcommand{\bea}{\begin{eqnarray}}
\newcommand{\eea}{\end{eqnarray}}
\newcommand{\non}{\nonumber}

\graphicspath{{./}{figures/}}

\received{}
\revised{}
\accepted{}

\shorttitle{Testing alternative theories of gravity}
\shortauthors{Shahzadi et al.}

\begin{document}

\title{Testing alternative theories of gravity by fitting the hot-spot data of Sgr~A*}

\correspondingauthor{Misbah Shahzadi}
\email{misbahshahzadi51@gmail.com}

\author[0000-0002-3130-1602]{Misbah Shahzadi}
\affil{Department of Mathematics, COMSATS University Islamabad, Lahore Campus, 54000 Lahore, Pakistan}

\author[0000-0002-4900-5537]{Martin Kolo{\v s}}
\affil{Research Centre for Theoretical Physics and Astrophysics, Institute of Physics, \\Silesian University in Opava, Bezru\v{c}ovo n\'{a}m.13, CZ-74601 Opava, Czech Republic}

\author[0000-0003-2178-3588]{Zden{\v e}k Stuchl\'ik}
\affil{Research Centre for Theoretical Physics and Astrophysics, Institute of Physics, \\Silesian University in Opava, Bezru\v{c}ovo n\'{a}m.13, CZ-74601 Opava, Czech Republic}

\author[0000-0003-1709-1159]{Yousaf Habib}
\affil{Department of Mathematics, COMSATS University Islamabad, Lahore Campus, 54000 Lahore, Pakistan}

\begin{abstract}
We are fitting dynamics of electrically neutral hot-spot orbiting around Sgr~A* source in Galactic center, represented by various modifications of the standard Kerr black hole (BH), to the three flares observed by the GRAVITY instrument on May~27, July~22, July~28, 2018. We consider stationary, axisymmetric and asymptotically flat spacetimes describing charged BHs in general relativity (GR) combined with non-linear electrodynamics, or reflecting influence of dark matter (DM), or in so called parameterized dirty Kerr spacetimes. We distinguish the spacetimes having different orbital frequencies from the standard Kerr BH, and test various BH spacetimes using the hot-spot data. We show that the orbital frequencies and positions of the hot-spots orbiting the considered BHs, fit the observed positions and periods of the flare orbits and give relevant constrains on the parameters of the considered BH spacetimes and the gravity or other theories behind such modified spacetimes.
%
\end{abstract}


\tableofcontents

\section{Introduction} \label{sec:intro}

General Relativity is an elegant theory of gravity that agrees with all observations at Solar System scale and beyond \citep{Will:2001:LRR}. The most famous results of GR are: the bending of light due to the gravitational field \citep{Will:2015:CQGra}, the correction of precession of perihelion of Mercury \citep{Will:2018:PhRvL}, and the existence of gravitational waves \citep{Abbott-et-al:2016:PhRvL}. Another important prediction of GR is the existence of astrophysical objects (BHs) with strong gravitational interaction from where nothing can escape, even the light. However, the  nonlinear behaviour and strong-field structure of GR still remains elusive and difficult to test \citep{Psa:2008:LRR}.

Observational data based on the dynamics of whole Universe affirm that the major part of mass of the universe is invisible (in modern cosmology, this invisible mass is known as DM). Even larger part of invisible content of the Universe is related to the so called dark energy (DE) \citep{Cal-Kam:2009:Natur} that can be well represented by the cosmological constant. For its relevance in astrophysical processes, see \citep{Stu-Hle:1999:PRD,Stuchlik:2005:MPLA,Sla-Stu:2005:CQGra,Bal-Mot-Now:2007:MNRAS,Stu-etal:2020:Universe:}.

Modern cosmological observations also reveal that our Universe is composed of $68.3 \%$ DE, $26.8 \%$ DM, and $4.9 \%$ ordinary matter \citep{Plank-Coll:2014:A&A,Rez:2017:ApJ}. Dark matter surrounding the galaxies and clusters does not interact with baryonic matter but can be observed by its gravitational effects on visible matter. Babcock \citep{Bab:1939:LicOB} examined the rotational speed of luminous objects in Andromeda galaxy and found that rotational speed continuously increases as one moves away from the center of these objects. This demonstrates that outer region of that luminous part is dominated by matter which does not shine. Zwicky  \citep{Zwicky:2009:GReGr} found a large amount of unseen (non-luminous) matter in the Universe rather than the seen (luminous) and detected the non-luminous mass lying outside the luminous parts of the galaxies. Besides of these theoretical observations, there is no experimental success in detecting DM yet.

In addition to the need for DE and DM, in the study of BHs, a problem that appears in GR, is the presence of singularities that are points or set of points where the geodesic is interrupted and the physical quantities diverge \citep{Haw-Pen:1970:PRSLSA,Bro-Rub:2013:bhce:conf}. It is believed that the problem of singularity occurs because the theory is classical and that in the quantum theory of gravity this problem would be solved. This, together with some long-standing problems in GR (like difficulties in explaining the accelerated Universe and galaxy rotation curves, etc), has motivated the study of viable alternative theories of gravity. These theories, also known as modified theories of gravity, aim to reproduce GR in the weak-field regime, but they can differ substantially from it in the strong curvature regime, where non-linear effects become dominant. These modified theories of gravity are developed by modifying the matter or gravitational part of the Einstein-Hilbert action.

Astronomers have classified the astrophysical BH candidates into three major classes (depending on the mass of BHs): stellar-mass BHs with mass $M \sim 5 M_{\odot} - 20 M_{\odot} $ situated in X-ray binary systems; super-massive BHs having $M \sim 10^{5} M_{\odot} - 10^{9} M_{\odot} $ found in galactic nuclei; and intermediate-mass BHs with $M \sim 10^{2} M_{\odot} - 10^{4} M_{\odot}$ \citep{Narayan:2005:NJPh}. The third class of objects (intermediate-mass BHs) is still debatable because their observations are indirect and dynamical measurements of their masses are still lacking.

The Sagittarius A* (Sgr~A*) is very compact and bright astronomical radio source situated at the Galactic Center of Milky Way, associated with the supermassive BH, and considered as a highly variable source across all the wavelengths \citep{Eck-Sch-Str:2005:bhcm:book,Melia:2007:gsbh:book}. A precise measurement of its spin and mass is a long standing problem for astrophysicists. The mass of the supermassive BH Sgr~A* estimated by the observed orbital parameters of the S0 star traveling in the BH's gravitational field with velocity of $\sim 10^{3} km/s$ is $M = (4.1 \pm 0.4) \times 10^{6} M_{\odot}$ \citep{Gil-et-al:2009:ApJ}. The mass and spin of Sgr~A* have been estimated using different approaches in several studies. \citep{Dok:2014:GReGr, Dok:2016:gac:conf} has been estimated the values of mass $M = (4.2 \pm 0.2) \times 10^{6} M_{\odot}$, and spin $a= 0.65 \pm 0.05$ with the help of the observed quasi-periodic oscillations from the supermassive BH in the Galactic center in X-rays.
However, the observed high-frequency QPOs could indicate a much higher spin on $a\sim0.999$ and relevance of the so called Aschenbach effect \citep{Asch:2004:AAP,Stu-etal:2005:PRD}

Near-infrared GRAVITY@ESO observations \citep{Grav-Coll:2018:A&A} have revealed the detection of two bright flares on July 22 and July 28, 2018 as well as a fainter flare on May 27, 2018, in the background of Galactic center massive BH. These flares are found with a peak approaching the flux of S2, and remain for 30–90 min.  The GRAVITY observations of hot-spots nearby the ISCO of Sgr A* BH allow us to fit their orbital radii as well as orbital periods with the circular orbits of test particles orbiting Kerr BH with mass $M \sim 4$ million  $M_{\odot}$.

The dynamics of charged hot-spot around supermassive Kerr BH in the presence of magnetic field as well as the equatorial circular orbits fitting the observed periods and positions of three flares have been explored by \cite{Tur-etal:2020:APJ:}. Recently, \citep{Mat-Cha-Pir:2020:MNRAS} investigated the July 22 flare in the background of several different models including geodesics, circular Keplerian, precessing pattern, and super-Keplerian pattern (a hot-spot moving faster than Keplerian along a circular trajectory), and fit the hot-spot trajectories. They proposed that a super-Keplerian circular orbits with orbital frequency $\Omega = 2.7 ~ \Omega_{k}$ at $r = 12.5 M $ yields a better match to the data than the Keplerian orbits, where $\Omega_{k}$ is the orbital frequency of Keplerian orbits. A model for the flares formulated on general relativistic magneto-hydrodynamic simulations of magnetically arrested accretion disks which present the violent episodes of flux escape from BH magnetosphere has been discussed by \cite{Por-Miz-You:2021:MNRAS}.

The nature and origin of hot-spots or flares still remains unclear. In addition, there is no observational information regarding shape of hot-spot. We assume that the hot-spot is a bound test, potentially neutral mass moving on a circular orbit in the background of Kerr spacetime, or in some of different BH spacetimes. We also assume that shape of the hot-spot as well as the surrounding environment does not change during one orbital timescale.

In this study, we consider the classical Kerr BHs, rotating regular BHs in GR (regular Bardeen, regular ABG), a well known Johannsen-Psaltis spacetime (non-Kerr BHs), and various other metrics in several theories of gravity, i.e., Kerr-Sen BHs in heterotic string theory, Born-Infeld BHs in Einstein-Born-Infeld theory, Kalb-Ramond BHs in heterotic string theory, Gauss-Bonnet BHs in Einstein-Gauss-Bonnet theory, Konoplya-Zhidenko BHs in an unknown alternative theory of gravity, rotating BHs in perfect fluid DM, as well as rotating regular BHs in Einstein-Yang-Mills theory, and explore the dynamics of neutral hot-spot orbiting the considering BHs. We calculate the corresponding orbital frequency, and examine the equatorial circular orbits of neutral hot-spot fitting the observed positions and periods of the three flares proposed by GRAVITY on July 22, May 27, and Jul 28, 2018. From dependence of the fitting lines on the parameters of various BH spacetimes we immediately indicate spacetime promising good fits for the data of the three flares.

Throughout the paper, we use the space-like signature ($-,+,+,+$) and the system of units in which $c = 1$ and $G = 1$. However, for the expressions with an astrophysical application and estimates, we use the units with the gravitational constant and the speed of light. Greek indices are taken to run from 0 to 3; Latin indices are related to the space components of the corresponding equations.

\section{Stationary and axi-symmetric Spacetimes} \label{SAS}

In four-dimensional GR, the no-hair theorem \citep{Israel:1967:PhRv,Carter:1971:PhRvL} states that the uncharged rotating BHs are uniquely characterized by only two parameters, the mass $M$, and spin $a$ of the BH, and are governed by the Kerr metric. This metric is a unique axisymmetric, stationary, asymptotically flat, and vacuum solution of the Einstein field equations which possesses an event horizon but there is no closed timelike curves in an exterior domain. Due to the weak cosmic censorship conjecture \citep{Penrose:1969:NCimR}, the central singularity is always behind the event horizon. However, the hypothesis that the astrophysical BH candidates are characterized by the Kerr spacetimes still lacks the direct evidence, furthermore, the GR has been tested only in the regime of weak gravity \citep{Will:2014:LRR}. For strong gravitational fields, the GR could be broken down and astrophysical BHs might not be the Kerr BHs as predicted by the no-hair theorem \citep{Joh-Psa:2011:PhRvD}. Several parametric deviations from the Kerr metric have been proposed to investigate the observational signatures in both the electromagnetic and gravitational-wave spectral that differ from the expected Kerr signals.

The line element of an arbitrary, stationary, axi-symmetric, and asymptotically flat spacetime with refelction symmetry reads
\beq
d s^2 = g_{tt}d{t}^2 + g_{rr}d{r}^2 + g_{\theta\theta} d\theta^2 + g_{\phi\phi} d\phi^2 + 2g_{t\phi} d{t} d \phi, \label{Kerr-likeMetric}
\eeq
where metric components $g_{\alpha \beta}$ are functions of $r$, $\theta$, and some additional parameters. In the following, we consider several stationary, axisymmetric, and asymptotically flat spacetimes both in GR and modified theories of gravity.

\subsection{Classical BHs (Kerr) in GR}

The nonzero components of the metric tensor $g_{\mu\nu}$, describing the geometry of the well known classical neutral rotating Kerr BH, taking in the standard Boyer-Lindquist coordinates can be written in the form \citep{Kerr:1963:PRL:,Carter:1968:PR:}
\bea
g_{tt} &=& -\left(\frac{\Delta-a^{2}\sin^{2}\theta}{\RS}\right), \quad
g_{rr} = \frac{\RS}{\Delta}, \quad g_{\theta\theta} = \RS,\nonumber\\
g_{\phi\phi} &=& \frac{\sin^{2}\theta}{\RS}\left[(r^{2}+a^{2})^{2}-\Delta
a^{2}\sin^{2}\theta\right], \nonumber\\
g_{t\phi}&=&\frac{a\sin^{2}\theta}{\RS}\left[\Delta-(r^{2}+a^{2})\right] ,
\label{MetricCoef}
\eea
with
\bea
\Delta &=& r^2 - 2Mr + a^2, \\ \RS &=& r^2 + a^2 \cos^2\theta, \label{KerrSigma}
\eea
where, $M$ and $a$ are the mass and rotation parameter of the BH, respectively. The spin parameter $a$ is bounded by $a \leq M$. The horizons for Kerr BH can be found by solving the condition $\Delta = 0$.

\subsection{Charged BHs in GR}

According to the no-hair theorem, BH solutions of the Einstein-Maxwell equations of GR (combining the field equations of gravity and electromagnetism) are fully characterised by their mass $M$, rotation parameter $a$, and electric charge $Q$. There are many kinds of charges such as electric, magnetic, tidal, and dyonic, etc. In the following we consider BH solutions with different charges.

\subsubsection{Kerr-Newmann BHs}

The non-zero components of metric co-effiecients of Kerr-Newman (KN) BH takes the form \citep{Mis-Tho-Whe:1973:Gra:,Bic-Stu-Bal:1989:BAC:}
\bea
g_{tt} &=& -\left(\frac{\Delta_{\rm q} - a^{2}\sin^{2}\theta}{\RS}\right), \quad
g_{rr} = \frac{\RS}{\Delta_{\rm q}}, \quad g_{\theta\theta} = \RS,\nonumber\\
g_{\phi\phi} &=& \frac{\sin^{2}\theta}{\RS}\left[(r^{2} + a^{2})^{2} - \Delta_{\rm q} ~
a^{2}\sin^{2}\theta \right], \nonumber\\
g_{t\phi}&=&\frac{a\sin^{2}\theta}{\RS}\left[\Delta_{\rm q} - (r^{2}+a^{2})\right],
\eea
with
\beq
\Delta_{\rm q} = r^2 - 2Mr + a^2 + \tilde{Q},
\eeq
where $\tilde{Q}$ is the electric charge of KN BH. For vanishing charge ($\tilde{Q} = 0$), the KN BHs reduces to the Kerr BH solutions.

\subsubsection{Braneworld BHs}

Rotating charged BHs in the brany universe of the Randall–Sundrum type with infinite additional
dimension are described by the Kerr geometry with an additional parameter, represented by the line element with following metric co-efficients \citep{Ali-Gur:2005:PhRvD:,Kot-Stu-Tor:2008:CLAQG:,Stu-Kot:2009:GReGr:}
\bea
g_{tt} &=& -\left(\frac{\Delta_{\rm b} - a^{2}\sin^{2}\theta}{\RS}\right), \quad
g_{rr} = \frac{\RS}{\Delta_{\rm b}}, \quad g_{\theta\theta} = \RS,\nonumber\\
g_{\phi\phi} &=& \frac{\sin^{2}\theta}{\RS}\left[(r^{2}+a^{2})^{2} - a^2 ~ \Delta_{\rm b} \sin^{2}\theta\right], \nonumber\\
g_{t\phi}&=&\frac{a\sin^{2}\theta}{\RS}\left[\Delta_{\rm b}-(r^{2} + a^{2})\right],
\eea
with
\bea
\Delta_{\rm b} &=& r^2 - 2Mr + a^2 + \beta.
\eea
The tidal charge parameter $\beta$ represents the interaction between brany BH and bulk spacetime, and can be both negative and positive. The negative tidal charge can provide a mechanism for spinning up the BH so that its rotation parameter exceeds its mass, which is not allowed in the framework of GR.

\subsubsection{Dyonic charged BHs}

A particle having both electric and magnetic charge is called dyon. The possibility of existence of dyonic
BHs is either due to magnetic monopoles raised into grand unification theories or it may be primordial. The metric co-efficients for rotating charged dyonic BHs are given by \citep{Kasuya:1982:PhRvD:,Stu:1983:BAC:}
\bea
g_{tt} &=& -\left(\frac{\Delta_{\rm d} - a^{2}\sin^{2}\theta}{\RS}\right), \quad
g_{rr} = \frac{\RS}{\Delta_{\rm d}}, \quad g_{\theta\theta} = \RS,\nonumber\\
g_{\phi\phi} &=& \frac{\sin^{2}\theta}{\RS}\left[(r^{2}+a^{2})^{2} - a^2 ~ \Delta_{\rm d} \sin^{2}\theta\right], \nonumber\\
g_{t\phi}&=& -\frac{a\sin^{2}\theta}{\RS}\left[(r^{2} + a^{2}) - \Delta_{\rm d}\right],
\eea
with
\bea
\Delta_{\rm d} &=& r^2 - 2Mr + a^2 + Q_{e}^2 + Q_{m}^2,
\eea
where $Q_{e}$ is electric charge, and $Q_{m}$ is the magnetic charge.

\subsubsection{Kerr-Taub-Nut BHs}

The Kerr-Taub-Nut (Newman-unti-tamburino) solution is an analytic type D vacuum solution of the Einstein equations and can be represented by \citep{Dem-New:1966:BAPSS:,Miller:1973:JMP:}
\bea
g_{tt} &=& -\left(\frac{\Delta_{\rm u} - a^{2}\sin^{2}\theta}{\RS_{\rm u}}\right), \quad
g_{rr} = \frac{\RS_{\rm u}}{\Delta_{\rm u}}, \quad g_{\theta\theta} = \RS_{\rm u},\nonumber\\
g_{\phi\phi} &=& \frac{1}{\RS_{\rm u}} \left[(\RS_{\rm u} + a ~ \tilde{\chi})^2 \sin^{2} \theta - \tilde{\chi}^{2} \Delta_{\rm u} \right], \nonumber\\
g_{t\phi}&=& \frac{2}{\RS_{\rm u}}\left[\Delta_{\rm u} ~ \tilde{\chi} - a (\RS_{\rm u} + a ~ \tilde{\chi}) \sin^{2} \theta \right],
\label{MetricCoef-KT}
\eea
with
\bea\nonumber
\tilde{\chi} &=& a \sin^{2} \theta - 2 n \cos \theta, \quad \RS_{\rm u} = r^2 +(n + a \cos \theta)^2,\\
\Delta_{\rm u} &=& r^2 - 2Mr + a^2 - n^2,
\eea
where $n$ is the gravitomagnetic or Nut charge, and for limiting value $n=0$, Eq. (\ref{MetricCoef-KT}) reduces to the usual Kerr BH solution. The Kerr-Taub-Nut spacetime is asymptotically non-flat due to the Nut charge, and there are string singularities on the symmetric axis.

\subsubsection{KN-Taub-Nut BHs}

The KN-Taub-Nut BH is stationary and axisymmetric non-vacuum object, completely described by mass, rotation, an electric charge, and nut parameter, written in the form \citep{Dem-New:1966:BAPSS:,Miller:1973:JMP:}
\bea
g_{tt} &=& -\left(\frac{\Delta_{\rm k} - a^{2}\sin^{2}\theta}{\RS_{\rm u}}\right), \quad
g_{rr} = \frac{\RS_{\rm u}}{\Delta_{\rm k}}, \nonumber\\
g_{\theta\theta} &=& \RS_{\rm u}, \quad
g_{\phi\phi} = \frac{1}{\RS_{\rm u}} \left[(\RS_{\rm u} + a ~ \tilde{\chi})^2 \sin^{2} \theta - \tilde{\chi}^{2} \Delta_{\rm u} \right], \nonumber\\
g_{t\phi}&=& \frac{2}{\RS_{\rm u}}\left[\Delta_{\rm k} ~ \tilde{\chi} - a (\RS_{\rm u} + a ~ \tilde{\chi}) \sin^{2} \theta \right],
\label{MetricCoef-KNT}
\eea
with
\bea
\Delta_{\rm k} = r^2 - 2Mr + a^2 + Q_{n}^{2} - \tilde{n}^2,
\eea
where $Q_{\rm n}$ is the electric charge, and $\tilde{n}$ is the Nut charge parameter of KN-Taubt-Nut BH. For $Q_{\rm n} = 0$, Eq. (\ref{MetricCoef-KNT}) reduces to the Kerr-Nut BH, and $Q_{\rm n} = 0 = \tilde{n}$ leads to the Kerr BH.

\subsection{Bumpy spacetimes in GR}

It is possible that the spacetime around massive compact objects which are assumed to be BH is not described by the Kerr metric, but by a metric which can be considered as a perturbation of the Kerr metric, and are usually known as bumpy (non-Kerr) spacetimes \citep{Col-Hug:2004:PhRvD:}. These spacetimes have multipoles and possesses some features that deviate slightly from the Kerr spacetime, reducing to the classical Kerr BH solutions when the deviation is zero. Here, we consider some bumpy spacetimes in GR.

\subsubsection{Johannsen-Psaltis spacetime}

In order to test the gravity in the region of strong gravitational field, Johannsen and Psaltis \citep{Joh-Psa:2011:PhRvD} proposed a deformed Kerr-like metric which describes the geometry of a stationary, axi-symmetric, and asymptotically flat vacuum spacetime, and the corresponding non-zero metric co-efficients can be written in the form
\bea\non
g_{tt} &=& -\left(1 - \frac{2 M r}{\RS}\right) \left(1 + h(r) \right),\\
g_{rr} &=& \frac{\RS (1+ h(r))}{\Delta + h(r) a^{2}\sin^{2} \theta}, \quad g_{\theta\theta} = \RS,\nonumber\\
g_{\phi\phi} &=& \sin^{2}\theta \left[\RS + \frac{a^{2}(2Mr + \RS)\sin^{2} \theta}{\RS} (1+h(r))  \right], \nonumber\\
g_{t\phi}&=& - \frac{2 a M )}{\RS}(1+h(r)) \sin^{2}\theta ,
\label{MetricCoef-Non-Kerr}
\eea
where
\beq
h(r) = \frac{M^{3} r \epsilon}{\RS^{2}}.
\eeq
The deformation parameter $\epsilon$ determines the degree of variation that the BH is more oblate $(\epsilon < 0)$, or prolate ($\epsilon > 0$) than the Kerr BH, and one can restore the Kerr metric for limiting case $\epsilon = 0$. The Johannsen-Psaltis metric is a perturbation of the Kerr metric designed to avoid pathologies like naked singularities and closed timelike curves.

\subsubsection{Hartle-Thorne spacetime}

The other solution that deals with quadrupole to linear, and rotation to second order is the
Hartle-Thorne metric. It is an approximate solution of GR equations and can describe an inner source for the compact object. The corresponding metric co-efficients read \citep{Har-Tho:1968:ApJ:}
\bea\non
g_{tt} &=& - F_{1}, \quad g_{rr} = \frac{1}{F_{2}}, \quad g_{\theta\theta} = r^{2} F_{3},\nonumber\\
g_{\phi\phi} &=& r^2 F_{3} \sin^{2}\theta, \quad  g_{t\phi}= - \frac{2 a M }{r} \sin^{2}\theta
,
\label{Hartle-Thorne}
\eea
with
\bea
\mathbb{F} &=& 1-\frac{2 M}{r} + \frac{2 a^2}{r^4}, \quad \mathbb{\tilde{F}} = \frac{2 M \mathcal{Q}_{2}^{1}(x)}{\sqrt{r(r-2M)}}  - \mathcal{Q}_{2}^{2}(x), \nonumber\\
F_{1} &=& \mathbb{F}\left[1 + \frac{2 a^2}{M r^3} \left(1+\frac{M}{r}\right) P_{2} (y) + 2 \mathbb{Q} \mathcal{Q}_{2}^{2} (x) P_{2} (y)\right], \nonumber\\\
F_{2} &=& \mathbb{F}\left[1 + \frac{2 a^2}{M r^3} \left(1 - \frac{5 M}{r}\right) P_{2} (y) + 2 \mathbb{Q} \mathcal{Q}_{2}^{2} (x) P_{2} (y)\right], \nonumber\\
F_{3}&=& 1 - \frac{2 a^2}{M r^3} \left(1+ \frac{2M}{r}\right) P_{2}(y) + 2 \mathbb{Q} \mathbb{\tilde{F}} P_{2}(y),
\eea
and
\beq
\mathbb{Q} = q_{1} - \frac{5 a^{2}}{8 M^4}, \quad x = -1 + \frac{r}{M}, \quad y = \cos\theta,
\eeq
where, $q_{1}$ is the quadrupole parameter. The expressions for the Legendre functions $\mathcal{Q}_{2}^{1} (x), \mathcal{Q}_{2}^{2} (x)$ of the second kind in interval $x \in [1, \infty]$, and $P_{2}(y)$ can be found in \citep{Har-Tho:1968:ApJ:}.

\subsubsection{Kerr-Q spacetime}

The Kerr-Q metric is the simplest extension of Kerr metric that admit a quadruple and has been tested to be singularity free outside the horizon. It is derived from the rotating $\delta$-metric using $\delta = (1 + q_{0})$, $m = M/(1 + q_{0})$ and expanding to the first order in $q_{0}$, second order in rotation parameter $a$, where $M$ is the physical mass. The corresponding metric co-efficients read \citep{All-Fir-Mas:2020:CQGra:}
\bea\non
g_{tt} &=& -\left(\mathbb{A} + q_{0}\left(\frac{2 M}{r\mathbb{A}} + \ln \mathbb{A}\right)\mathbb{A} + \frac{2 a^2 M}{r^3} \cos^{2} \theta \right),\\
g_{rr} &=& \frac{1}{\mathbb{A}}- \frac{q}{\mathbb{A}} \left(\frac{2M}{r \mathbb{A}} + \ln \frac{\mathbb{B}^2}{\mathbb{A}}\right) - \frac{a^2 \mathbb{A}}{r^2 \mathbb{A}} (1- \cos^{2} \theta), \nonumber\\
g_{\theta\theta} &=& \left(1- q_{0} \ln \frac{\mathbb{B}^2}{\mathbb{A}} + \frac{a^2}{r^2}\cos^{2}\theta \right)r^2,\nonumber\\
g_{\phi\phi} &=& r^2 \sin^{2}\theta \left[1 - q_{0} \ln \mathbb{A} + \frac{a^{2}}{r^2} \left(1 + \frac{2 M}{r} \sin^{2} \theta \right)  \right], \nonumber\\
g_{t\phi}&=& - \frac{2 a M }{r} \sin^{2}\theta ,
\label{Kerr-Q}
\eea
where
\beq
\mathbb{A} = 1-\frac{2 M}{r}, \quad \mathbb{B} =  1-\frac{2 M}{r} + \frac{M^2}{r^2} \sin^{2} \theta.
\eeq
Here, $q_{0}$ is the quadrupole parameter that determines the deviations from the Kerr BH, and for $q_{0} = 0$, Eq. (\ref{Kerr-Q}) reduces to the Kerr BH.

\subsubsection{Quasi-Kerr BHs}

The general stationary axisymmetric neutral compact object can bee characterized by mass, multipole moments, and rotational parameter.  The multipole moments are consist of a set of mass multipole moment $M_{\rm k}$ and current multipole moment $S_{\rm k}$ , here the subscript $k$ of them is labeled by the angular inter eigenvalue $k \geq 0$. The relation between the parameters of the multipole moments can be expressed as
\beq
M_{\rm k} + i S_{\rm k} = M (ia)^{\rm k} + \delta M_{\rm k} + i \delta S_{\rm k}.
\eeq
For classical Kerr BHs, the deviation $\delta M_{\rm k}$ and $\delta S_{\rm k}$ vanishes, while the BH solutions with the quadrupole moment takes the form
\beq
g_{\alpha \beta} = g^{\rm Kerr}_{\alpha \beta} + \tilde{\epsilon} ~ h_{\alpha \beta},
\eeq
where $g^{\rm Kerr}_{\alpha \beta}$ indicates the metric tensor for classical Kerr BHs, and the components of $h_{\alpha \beta}$ can be written as \citep{Gla-Kos-Sta:2006:CQGra:}
\bea
\mathcal{Y} &=& 1- 3 \cos^{2} \theta, \quad h^{tt} = \left(1 - \frac{2 M}{r} \right)^{-1} \mathcal{Y} \mathcal{F}_{1}(r), \nonumber\\
h^{rr} &=& \left(1 - \frac{2 M}{r} \right)\mathcal{Y} \mathcal{F}_{1}(r), \quad
h^{\theta \theta} = - \frac{\mathcal{Y} \mathcal{F}_{2}(r)}{r^2}, \nonumber\\
 h^{\phi \phi} &=& - \frac{\mathcal{Y} \mathcal{F}_{2}(r)}{r^2 \sin^{2}\theta}, \quad
 h^{t\phi} = 0,
\eea
with the functions $\mathcal{F}_{1,2}(r)$ shown explicitly in Appendix A of \citep{Gla-Kos-Sta:2006:CQGra:}. The deformation parameter $\tilde{\epsilon}$ indicates a small contribution to the quadrupole moment $q$ of the compact object with the total mass $M$ as
\beq
q = -M (a^2 + \tilde{\epsilon} ~ M^2),
\eeq
and can take both negative and positive values. For vanishing $\tilde{\epsilon} = 0$, Quasi-Kerr BHs reduce to the classical Kerr BH solutions.

\subsubsection{Accelerating and rotating BHs}

The accelerating and rotating BH solutions describe the gravitational field by a pair of uniformly
accelerating Kerr-type BHs, which is a special case of the Pleba\'nski and Demia\'nski spacetime that covered a large family of electro-vacuum type-D spacetimes including both the KN like solutions and the C-metric, the corresponding metric co-efficients read \citep{Gri-Pod:2005:CQG:}
\bea
g_{tt} &=& -\left(\frac{\Delta_{\rm A} - a^{2} \mathcal{P} \sin^{2}\theta}{\RS \mathcal{P}_{1}^{2}}\right), \quad
g_{rr} = \frac{\RS}{\Delta_{\rm A} \mathcal{P}_{1}^{2}}, \nonumber\\
g_{\theta\theta} &=& \frac{\RS}{\mathcal{P} \mathcal{P}_{1}^{2}}, \quad
g_{\phi\phi} = \frac{\sin^{2}\theta}{\RS \mathcal{P}_{1}^{2}}\left[\mathcal{P}(r^{2}+a^{2})^{2}-\Delta_{\rm A}
a^{2}\sin^{2}\theta\right], \nonumber\\
g_{t\phi}&=&\frac{-2 a\sin^{2}\theta}{\mathcal{P}_{1}^{2} \RS}\left[\mathcal{P}_{1}^{2} (r^{2} + a^{2})- \Delta_{\rm A}\right] ,
\label{Acc-Rot}
\eea
with
\bea
\mathcal{P} &=& 1- 2 M \tilde{b} \cos \theta +  \tilde{b}^{2} a^{2} \cos \theta, \quad
\mathcal{P}_{1} = 1- r \cos \theta, \nonumber\\
\Delta_{\rm A} &=& (r^2 - 2Mr + a^2)(1- \tilde{b}^{2} r^{2}).
\eea
The parameter $\tilde{b}$ determines the acceleration of the BH. The accelerating and rotating BHs have the same event and Cauchy horizons as the Kerr BH, but there also exits two other horizons which can be interpreted as the acceleration horizons, i.e.,
\beq
r_{\pm} = M \pm \sqrt{M^2 - a^2}, \quad r_{\rm A} = \frac{1}{\tilde{b}}, \quad r_{\rm B} = \frac{1}{\tilde{b} \cos \theta}.
\eeq
Thus, unlike in the usual Kerr BH spacetime, the physical region of this BH is situated in $ r_{+} < r < r_{\rm A}$, where $\Delta_{\rm A} > 0$ is satisfied.

\subsection{Rotating regular BHs in GR}

The regular BHs are non-singular exact solutions of the Einstein field equations minimally coupled to a non-linear electrodynamics, satisfy the weak energy condition, and yield alteration to the classical BHs \citep{Stu-Sch:2015:IJMPD}. The regular BHs are constructed to be regular everywhere, i.e., the Ricci scalar, and the components of the Riemann tensor are finite $\forall ~ r \geq 0$. There are three well known regular BHs in GR, Bardeen rotating regular BHs, and Ay\'{o}n-Beato-Garica (ABG) regular BHs, and Hayward regular BHs \citep{Hayward:2006:PRL:,Bec-et-al:2021:PhRvD}, which we describe below.

\subsubsection{Regular Bardeen BHs}

The spacetime filled with a vacuum can give a proper discrimination at the final stage of gravitational collapse, replacing the future singularity. Based on this idea, Bardeen \citep{Bardeen:1968:Tbilisi} proposed the first regular BH solution named as Bardeen regular BH, according to whom there is no singularity but horizons can exist \citep{Tos-etal:2014:PRD:,Stu-Sch:2019:EPJC}. The matter field is a kind of magnetic field and the solution yields a modification of the classical Kerr BH solution. The non-zero metric co-efficients corresponding to the Bardeen regular BH read
\bea \label{Bardeen}
g_{tt} &=& -\left(1 - \frac{2 r m(r)}{\RS}\right), \quad
g_{rr} = \frac{\RS}{\Delta_{\rm B}}, \quad g_{\theta\theta} = \RS,\nonumber\\
g_{\phi\phi} &=& \sin^{2}\theta \left[r^{2}+a^{2} + \frac{2 r a^{2} m(r)}{\RS} \sin^{2} \theta \right], \nonumber\\
g_{t\phi}&=& - \frac{2 a r m(r)}{\RS}\sin^{2}\theta,
\label{MetricCoef-Bardeen}
\eea
where $\RS$ is defined by (\ref{KerrSigma}) and
\beq
\Delta_{\rm B} = r^2 - 2 r m(r) + a^2,
\eeq
and the mass function $m(r)$ takes the form
\beq
m(r) = M \left(\frac{r^2}{r^2 + q^2} \right)^\frac{3}{2}.
\eeq
The deviation parameter $q$ can be recognized as a magnetic monopole charge of non-linear electrodynamics, which determines the deviation from the Kerr BH, and when we turn-off the non-linear electrodynamics $(q=0)$, one can recover the Kerr metric, and for vanishing spin ($a=0$), we obtain the non-rotating Bardeen regular BH.

\subsubsection{Regular ABG BHs}

Another class of spherically symmetric static regular BH solutions was introduced by Ay\'on-Beato and Garc\'ia \citep{Ayo-Gar:1998:PhRvL}, and the rotating one is investigated by \citep{Tos-Stu-Ahm:2017:PhRvD}. The non-zero metric co-efficients of rotating ABG regular BH can be written in the form
\bea \label{ABG}
g_{tt} &=& -g(r, \theta), \quad
g_{rr} = \frac{\RS}{\RS g(r, \theta) + a^{2}\sin^{2}\theta }, \nonumber\\
g_{t \phi} &=& -a^{2}\sin^{2}\theta \left( 1- g(r, \theta)\right), \quad g_{\theta\theta} = \RS, \nonumber\\
g_{\phi \phi}&=&  \left(\RS + a^{2} - g(r, \theta)\sin^{2}\theta \right)\sin^{2}\theta,
\label{MetricCoef-ABG}
\eea
with
\beq
g(r, \theta) = 1- \frac{2 M r \sqrt{\RS}}{(\RS + Q^{2})^\frac{3}{2}} + \frac{\RS Q^{2}}{(\RS + Q^{2})^{2}},
\eeq
where $Q$ is the electric charge of the regular ABG BH, for vanishing charge $(Q=0)$, we obtain the Kerr metric, and for $a=0$, Eq. (\ref{MetricCoef-ABG}) reduces to the case of non-rotating ABG BH.

\subsubsection{Regular Hayward BHs}

The regular Hayward BH solutions can be described by Eq. (\ref{MetricCoef-Bardeen}) with the mass function \citep{Hayward:2006:PRL:}
\beq
m(r)=\frac{M r^3}{r^{3} + g^{3}},
\eeq
where $g$ is the deviation parameter, and for limiting case $g \rightarrow 0$, one can recover the classical Kerr BH.

\subsection{Rotating BHs in alternative theories of gravity}

The late-time acceleration of the Universe is surely the most challenging problem in cosmology. Many cosmological observations indicate that the accelerated expansion of the Universe is due to the existence of mysterious form of energy known as DE. Modern astrophysical and cosmological models are also faced with two severe theoretical problems, that can be summarized as the DM (non or weakly interacting), and the DE problem. The two observations, namely, the mass discrepancy in galactic clusters, and the behavior of the galactic rotation curves, suggest the existence of a DM at galactic and extra-galactic scales. Recently, several modified theories of gravity have been proposed to address these two intriguing and exciting problems facing modern physics. These modified theories of gravity are constructed by modifying the gravitational or matter part of the Einstein-Hilbert action. In addition, both non-rotating as well as rotating BH solutions has been derived in these modified theories of gravity \citep{Sen:1992:PhRvL,Moffat:2015:EPJC:,Shahzadi-et-al:2019:PDU}. In the following, we consider the several rotating BH solutions in many different modified theories of gravities.

\subsubsection{Kerr-Sen BHs}

Sen \citep{Sen:1992:PhRvL} proposed a charged rotating BH solution to the equations of motion of the low-energy limit of the heterotic string theory, known as the Kerr-Sen solution. The metric coefficients of the corresponding line element can be written as
\bea
    g_{tt} &=& -\left(1- \frac{2Mr}{\RS_{\rm K}}\right), \quad
    g_{rr} = \frac{\RS_{\rm K}}{\Delta_{\rm K}}, \quad g_{\theta\theta} = \RS_{\rm K},\nonumber\\
    g_{\phi\phi} &=& \left[a^{2} + r\left(\frac{Q^{2}_{K}}{M} + r \right) + \frac{2 a^2 M r \sin^{2}\theta}{\RS_{\rm K}}\right] \sin^{2}\theta, \nonumber\\
    g_{t\phi}&=&- \frac{4 a M r \sin^{2}\theta }{\RS_{K}},
\label{MetricCoef-KS}
\eea
with
\bea
    \Delta_{\rm K} &=& r\left(\frac{Q^{2}_{K}}{M} + r\right) + a^2 - 2Mr, \nonumber\\
    \RS_{\rm K} &=& r\left(\frac{Q^{2}_{K}}{M} + r\right) + a^2 \cos^{2}\theta,
\eea
where $Q_{K}$ is the electric charge for Kerr-Sen BH, and the metric (\ref{MetricCoef-KS}) reduces to the Kerr metric for the limiting value $Q_{K} \rightarrow 0$. The spacetime represented by the Kerr-Sen metric is not vacuum, and correspond to the KN case in the Einstein-Maxwell theory. There exists two horizons for a non-extremal Kerr-Sen BH, and can be determined by the condition $ \Delta_{\rm K} = 0$ as
\beq
r_{\pm} = M - \frac{Q_{K}^{2}}{2 M} \pm \sqrt{\left( M - \frac{Q_{K}^{2}}{2 M} \right)^{2} - a^{2}},
\eeq
where $r_{-}$, and $r_{+}$ correspond to the inner and outer horizons of the BH respectively. The range of the parameter $Q_{K}$ is bounded by $ 0 \leq Q_{K} \leq \sqrt{2} M $. For extremal BH, both horizons coincide, and we have the condition $Q_{K} = 2 (M-a) M$.

\subsubsection{Einstein-Born-Infeld BHs}

The gravitational field of a stationary and axisymmetric compact object with mass $M$, spin $a$ and a non-linear electromagnetic source in the Einstein-Born-Infeld theory has been investigated by Julio Cirilo Lombardo \citep{Julio:2004:CQGra}, and the metric coefficients of the spacetime read
\bea\non
    g_{tt} &=& -\left(\frac{\Delta_{\rm BI} -a^{2}\sin^{2}\theta}{\RS}\right),\\\non
    g_{rr} &=& \frac{\RS}{\Delta_{\rm BI}}, \quad g_{\theta\theta} = \RS,\nonumber\\
    g_{\phi\phi} &=& \frac{\sin^{2}\theta}{\RS}\left[(r^{2} + a^{2})^{2} - \Delta_{\rm BI} ~
    a^{2}\sin^{2}\theta\right], \nonumber\\
    g_{t\phi}&=&\frac{a\sin^{2}\theta}{\RS}\left[\Delta_{\rm BI} -(r^{2}+a^{2})\right] ,
\label{MetricCoef-BI}
\eea
with
\bea
    \Delta_{\rm BI} &=& r^2 - 2GMr + a^2 + Q^{2}(r), \nonumber\\
    Q^{2}(r) &=& \frac{2\beta^{2} r^4}{3} \left(1 - \sqrt{1 + \eta^{2} (r)}\right)\\\non
    &+& \frac{4Q_{BI}^2}{3} F \left(\frac{1}{4}, \frac{1}{2}, \frac{5}{4}, \eta^{2} (r) \right),
\eea
where $F$ denotes the Gauss hypergeometric function, $Q_{BI}$ shows the electric charge for Born-Infeld BH, $\eta^{2} (r) = Q_{BI}^{2}/r^{4} \beta^{2}$, and $\beta$ is the Born-Infeld parameter. For the limiting case $\beta = 0$, the metric (\ref{MetricCoef-BI}) reduces to the Kerr metric while for $\beta \rightarrow \infty$ (or $Q(r) = Q \neq 0$), we obtain KN metric. The Born-Infled parameter $\beta$ can take any positive real value, and the charge $Q_{BI}$ is bounded by $ 0< Q_{BI} <1$. The metric Eq. (\ref{MetricCoef-BI}) has curvature singularity at the points, where $M = Q \neq 0$, and $\RS = 0$. In an equatorial plane, it corresponds to a ring with radius $a$, and  termed as a ring singularity. The properties of the rotating Einstein-Born-Infeld BH (\ref{MetricCoef-BI}) are similar to that of the GR counterpart KN BH. Like the KN BH, Einstein-Born-Infeld BH is singular at $\Delta_{\rm BI} = 0$, and it admits thee static limit surface, two horizons like surfaces, and the event horizon. The horizons can be obtained by solving $\Delta_{\rm BI} =0 $, which are different from the KN BH, for details, see \citep{Ata-Gho-Ahm:2016:EPJC}.

\subsubsection{Kalb-Ramond BHs}

The Kalb-Ramond field is considered as a self-interacting, second-rank antisymmetric tensor field in the heterotic string gravity. It can also be considered as a generalization of the electromagnetic potential with two indices, such that the gauge potential $A_{\alpha}$ is replaced by the second-rank antisymmetric tensor field $B_{\alpha \beta}$ associated with the gauge-invariant rank-3 antisymmetric field strength $H_{\mu \alpha \beta} = \partial_{[ \mu} B_{\alpha \beta ]} $ \citep{Kum-Gho-Wan:2020:PhRvD}. The stationay, axisymmetric, asymptotically solution of the modified field equations leads to the hairy BH solution
\bea\non
g_{tt} &=& -\left(\frac{\Delta_{\rm KR}-a^{2}\sin^{2}\theta}{\RS}\right),\\\non
g_{rr} &=& \frac{\RS}{\Delta_{\rm KR}}, \quad g_{\theta\theta} = \RS,\nonumber\\
g_{\phi\phi} &=& \left[ \RS + a^2 \sin^{2} \theta \left( 2 - \frac{\Delta_{\rm KR}-a^{2}\sin^{2}\theta}{\RS}\right) \right]\sin^{2} \theta, \nonumber\\
g_{t\phi}&=& - a\sin^{2}\theta \left( 1- \frac{\Delta_{\rm KR}-a^{2}\sin^{2}\theta}{\RS} \right) ,
\label{MetricCoef-KR}
\eea
with
\beq
\Delta_{\rm KR} = r^2 - 2GMr + a^2 +  r^{2\frac{(s-1)}{s}} \Gamma.
\eeq
The spontaneous Lorentz violating parameters $s$, and $\Gamma$ are related to the vacuum expectation value of the Kalb-Ramond field and the non-minimal coupling parameter. The free parameter $s$ also known as Kalb-Ramond parameter, determines the potential deviation from the Kerr metric, and generalizes the KN metric. For the limiting case $s=0$, Eq. (\ref{MetricCoef-KR}) reverts to the Kerr BH, and to the KN BH for $s=1$. The numerical solution of $\Delta_{\rm KR} = 0$ reveals that there exist only two positive real roots corresponding to the inner and outer horizons. Two distinct real positive roots infers the non-extremal BH, while no BH in the absence of real positive roots, i.e., no horizon exists.

\subsubsection{Einstein-Gauss-Bonnet BHs}

The uniqueness of the Einstein field equations is build on the Lovelock theorem which states that the GR with cosmological constant is the only theory of gravity in four-dimensional spacetime. However, the Einstein-Hilbert action is not unique in higher-dimensional spacetimes ($d>4$, $d$ represents the dimension of the spacetime), and the Einstein-Gauss-Bonnet theory of gravity is one of the interesting example in higher dimensions \citep{Tor-Shi:2008:PhRvD}. Recently, this theory has been proposed in lower dimensions by re-scaling the Gauss-Bonnet coupling parameter $\alpha \rightarrow \alpha / (d-4)$, in the limit $d \rightarrow 4$ \citep{Gla-Chu:2020:PRL:}. This theory has attracted much attention, and both non-rotating and rotating BHs have been proposed in this theory. The study of four-dimensional BH solutions presents a new approach to understand the Gauss-Bonnet gravity in low dimensions. The non-zero metric co-efficients of rotating Einstein-Gauss-Bonnet BH can the written in the form \citep{Kum-Gho:2020:JCAP}
\bea\non
g_{tt} &=& -\left(\frac{\Delta_{\rm G}-a^{2}\sin^{2}\theta}{\RS}\right),\\\non
g_{rr} &=& \frac{\RS}{\Delta_{\rm G}}, \quad g_{\theta\theta} = \RS,\nonumber\\
g_{\phi\phi} &=& \frac{\sin^{2}\theta}{\RS}\left[(r^{2}+a^{2})^{2}-\Delta_{\rm G}~
a^{2}\sin^{2}\theta\right], \nonumber\\
g_{t\phi}&=&\frac{a\sin^{2}\theta}{\RS}\left[\Delta_{\rm G}-(r^{2}+a^{2})\right] ,
\label{MetricCoef-GB}
\eea
with
\beq
\Delta_{\rm G} = r^2 + a^2 + \frac{r^{4}}{32 \alpha \pi} \left(1 -\sqrt{1 + \frac{128 M \alpha \pi}{r^3}} \right),
\eeq
where $\alpha$ is the Einstein Gauss Bonnet coupling constant. For vanishing spin ($a=0$), one can obtain the non-rotating Einstein-Gauss-Bonnet BH, and for $\alpha = 0$, Eq. (\ref{MetricCoef-GB}) leads to the classical Kerr BH. The values of Gauss Bonnet parameter $\alpha$ fall in the range $  \alpha / M^{2} \in [-8, 1]$, and $\alpha >1$ lead to the naked singularity.

\subsubsection{Konoplya-Zhidenko BHs}

Konoplya and Zhidenko \citep{Kon-Zhi:2016:PhLB} proposed a rotating non-Kerr BH beyond GR and make an estimate for the possible deviations from the Kerr solution with the data of GW 150914, which can be regarded as a vacuum solution of an unknown alternative theory of gravity. The deformation changes the relation between the position of event horizon and BH mass, but preserves the asymptotic properties of Kerr spacetime. The non-zero metric co-efficients of the corresponding spacetime can be written in the form
 \bea\non
    g_{tt} &=& - \left(1 -  \frac{\eta + 2 M r^2  }{r \RS}\right),\\\non
    g_{rr} &=& \frac{\RS}{\Delta_{\rm KZ}}, \quad g_{\theta\theta} = \RS,\nonumber\\
    g_{\phi\phi} &=&  \left[r^2 + a^2 + \left(\frac{\eta + 2 M r^2}{r \RS} \right) a^2 \sin^{2}\theta \right]\sin^{2}\theta, \nonumber\\
    g_{t\phi}&=& -   \frac{a \left(\eta + 2 M r^2 \right) }{r \RS} \sin^{2}\theta ,
\label{MetricCoef-KZ}
\eea
with
\beq
\Delta_{\rm KZ} = r^2 + a^2 -2 M r - \frac{\eta}{r}.
\eeq
The deformation parameter $\eta$ of Konoplya-Zhidenko BH describes the deviations from the Kerr metric, and for vanishing $\eta$, one can obtain the usual Kerr spacetime. The presence of deformation parameter extends the allowed range of the spin parameter $a$ and changes the geometry of BH in the strong field region.


\subsubsection{Kerr-MOG BHs}
The Kerr-MOG BHs are the stationary, axially symmetric and asymptotically flat solutions of field equations scalar-tensor-vector gravity which can be considered as another alternative to GR without DM in the present Universe, and can be described the line element with metric co-efficient \citep{Moffat:2015:EPJC:,Kol-Sha-Stu:2020:EPJC}
\bea
g_{tt} &=& -\left(\frac{\Delta_{\rm m} - a^{2}\sin^{2}\theta}{\RS}\right), \quad
g_{rr} = \frac{\RS}{\Delta_{\rm m}}, \quad g_{\theta\theta} = \RS,\nonumber\\
g_{\phi\phi} &=& \frac{\sin^{2}\theta}{\RS}\left[(r^{2}+a^{2})^{2} - \Delta_{\rm m}
a^{2}\sin^{2}\theta\right], \nonumber\\
g_{t\phi}&=&\frac{a\sin^{2}\theta}{\RS}\left[\Delta_{\rm m} - (r^{2}+a^{2})\right],
\label{MetricCoef-KerrMOG}
\eea
where
\bea
\Delta_{\rm m} = r^2 - 2GMr + a^2 + \alpha G_{\rm N} G M^2.
\eea
Here, $G = G_{\rm N}(1+\alpha)$ is the enhanced gravitational constant, $M$ is the mass of the BH, $G_{\rm N}$ is Newton's gravitational constant. The dimensionless parameter $\alpha$ determines the gravitational field strength. For $\alpha=0$ and $a=0$, the Kerr-MOG BH reduces to the Kerr and Schwarzschild-MOG BH, respectively. For details, see \citep{Sha-Sha:2017:EPJC:,Sha-Sha:2018:JETP}.

\subsubsection{Kaluza-Klein BHs}

Kaluza-Klein BHs are the exact solutions of five-dimensional Kaluza-Klein (Einstein-Maxwell) theory. The simplest version of this theory is to study the GR in five dimensions. These theories are of great interest in string theory
community because of their roles as low-energy approximations to string theory. Larsen \citep{Larsen:2000:NuPhB:} proposed the most general BH solutions to the Einstein theory in five-dimensions and later dimensionally reducing the solution to four dimensions. The corresponding BH solution in an equatorial plane can be written using the follwing metric co-efficients \citep{Larsen:2000:NuPhB:}
\bea
g_{tt} &=& -\frac{\mathbb{H}_{3}}{\rho^{2}}, \quad
g_{rr} = \frac{\rho^{2}}{\Delta_{k}}, \quad g_{\phi\phi} = \frac{- \mathbb{H}_{4}^2  + \rho^{4} \Delta_{k}}{\mathbb{H}_{3}  \rho^2}, \nonumber\\
g_{t\phi}&=& \frac{\mathbb{H}_{4}}{\rho^2}, \quad \rho^2 = \sqrt{\mathbb{H}_{1} \mathbb{H}_{2}}.
\label{MetricCoef-KK}
\eea
The unknown quantities takes the form
\bea\nonumber
\frac{\mathbb{H}_{1}}{M^2} &=& \frac{8 b (\gamma-2) (b-2)}{(\gamma + b)^3} + \frac{4 x (b - 2)}{\gamma + b} + x^2,\\\nonumber
\frac{\mathbb{H}_{2}}{M^2} &=& \frac{8 \gamma (\gamma -2) (b-2)}{(\gamma + b)^3} + \frac{4 x (\gamma - 2)}{\gamma +b} + x^2,\\\nonumber
\frac{\mathbb{H}_{3}}{M^2} &=& x^2 - \frac{8x}{\gamma + b}, \quad \frac{\Delta_{k}}{M^2} = \alpha^{2} + x^2 - \frac{8x}{\gamma + b}, \\
\frac{\mathbb{H}_{4}}{M^2} &=& \frac{2 \sqrt{\gamma b } [(\gamma + b)(\gamma b + 4) - 4 (\gamma - 2)(b - 2)] \alpha}{(\gamma + b)^3},
\eea
where $\alpha \equiv a/M$, $x \equiv r/M$, and $b, \gamma$ are the dimensionless free parameters, for detail, see \citep{Gha-et-al:2020:PhRvD:}.

\subsubsection{BHs with Weyl corrections}

The generalized Einstein-Maxwell theories have received a lot of attention recently because it contains
higher derivative interactions and carries more information about the electromagnetic field. One of the simple generalized electromagnetic theories is the electrodynamics with Weyl corrections which involves a coupling between the Weyl tensor and Maxwell field. In this theory, the Lagrangian density of the electromagnetic field is modified as
\beq
\mathcal{L} = -\frac{1}{4}\left( F_{\alpha \beta} F^{\alpha \beta} -4 \tilde{\alpha} C^{\alpha \beta \gamma \delta} F_{\alpha \beta} F_{\gamma \delta} \right),
\eeq
where $F^{\alpha \beta}$ is the electromagnetic tensor associated with the electromagnetic vector potential $A^{\alpha}$, and $C^{\alpha \beta \gamma \delta}$ is the Weyl tensor. The coefficient $\tilde{\alpha}$ is a coupling constant with dimensions of length squared. The metric co-efficients describing the rotating BH with Weyl corrections can be written as \citep{Chen-Jin:2014:PhRvD:}
\bea\nonumber
g_{tt} &=& - X(r, \theta), \quad g_{rr} = \frac{\varpi}{\Delta_{w}}, \quad g_{\theta\theta} = \varpi,\nonumber\\
g_{\phi\phi} &=& \sin^{2}\theta \left[\varpi + a^2 \left(2 - X(r, \theta) \right) \sin^{2} \theta \right], \nonumber\\
g_{t\phi}&=& - 2 a (1 - X(r, \theta)) \sin^{2}\theta,
\label{Weyl-cor}
\eea
with
\bea\nonumber
X(r, \theta) &=& 1- \frac{2 M r}{\RS} - \frac{\tilde{q}^2}{\RS} - \frac{4 \tilde{\alpha} \tilde{q}^2}{3 \RS^2}\left(1- \frac{50 M r - 26 \tilde{q}^2}{15 \RS} \right),\\
\Delta_{w} &=& X(r, \theta) + a^2 \sin^{2} \theta, \quad \varpi = \RS + \frac{4 \tilde{\alpha} \tilde{q}^2}{9 \RS},
\eea
where $\tilde{q}$ is the electric charge. For the limiting case $\tilde{q} = \tilde{\alpha} = 0$, Eq. (\ref{Weyl-cor}) reduces to the Kerr BH, and for $\tilde{\alpha} = 0$, one can obtain the KN BH.\\

\subsubsection{BHs in Rastall gravity}

The Rastall theory of gravity is the modified GR, in which the usual energy-momentum conservation
law ($T^{\alpha \beta}_{; \alpha} =0$) is generalized to $T^{\alpha \beta}_{; \alpha} = \lambda R^{, \beta}$, where $T_{\alpha \beta}$ is the energy-momentum tensor, $\lambda$ is the Rastall parameter which represents the level of the energy-momentum conservation law in gravity theory. The metric co-efficients for rotating BHs in Rastall gravity reads \citep{Xu-et-al:2018:EPJC:}
\bea
g_{tt} &=& -\left(1 - \frac{2 M r + N_{\rm s} ~ r^{\xi}}{\RS}\right), \quad
g_{rr} = \frac{\RS}{\Delta_{\rm t}}, \quad g_{\theta\theta} = \RS,\nonumber\\
g_{\phi\phi} &=& \sin^{2}\theta \left[r^{2}+a^{2} + a^2 \left(\frac{2 M r + N_{\rm s} ~ r^{\xi}}{\RS} \right) \sin^{2} \theta \right], \nonumber\\
g_{t\phi}&=& - 2 a \left( \frac{ M r + N_{\rm s} ~ r^{\xi}}{\RS} \right) \sin^{2}\theta,
\label{MetricCoef-Rastall}
\eea
with
\bea\nonumber
\Delta_{\rm t} &=& r^2 - 2 M r + a^2 - N_{\rm s} ~ r^{\xi},\\
\xi &=& \frac{1 - 3 \omega_{s}}{1-3(1+ \omega_s ) \psi},
\eea
where $N_{\rm s}$ is the surrounding fluid structure parameter, $\psi$ is the Rastall coupling parameter, and $\omega_{s}$ is the state parameter of surrounding fluid. For vanishing parameter $N_{\rm s}$, Eq. (\ref{MetricCoef-Rastall}) goes over to the usual Kerr BH. For the limiting case $\psi = 0$, and $ -1 < \omega_{s} < -1/3$, the metric shows the Kerr BHs surrounded by quintessence.

\subsubsection{Charged Weyl BHs}

In Weyl theory of gravity, the Einstein-Hilbert action is modified by a term proportional to the square of the Weyl tensor, and the action of the standard model of particle physics is modified to make it be conformally invariant. This theory has been appeared as one of the alternatives to compare the theoretical results with the cosmological as well as astrophysical observations, and also describe the cosmological parameters relevant to DE problem. The rotating BH solutions in this theory leads to the charged Weyl BH solutions \citep{FAT-Oli-Vil:2021:Galax:}
\bea\non
g_{tt} &=& -\left(\frac{\Delta_{\rm c} -a^{2}\sin^{2}\theta}{\RS}\right),\quad g_{rr} = \frac{\RS}{\Delta_{c}}, \quad g_{\theta\theta} = \RS,\nonumber\\
   g_{\phi\phi} &=& \sin^{2}\theta \left[\RS +  \left(2- \frac{\Delta_{\rm c} - a^{2}\sin^{2}\theta}{\RS}\right )a^{2}\sin^{2}\theta \right], \nonumber\\
    g_{t\phi}&=& - 2a \left(1- \frac{\Delta_{\rm c} -a^{2}\sin^{2}\theta}{\RS}\right)\sin^{2}\theta ,
\label{MetricCoef-CW}
\eea
with
\bea\nonumber
\Delta_{\rm c} &=& r^2 + a^2 - \frac{r^4}{\lambda^2} - \frac{Q^{2}_{w}}{4},\\\
\frac{1}{\lambda^2} &=& \frac{3 \tilde{m} }{\tilde{r}^3} + \frac{2 \tilde{\epsilon}}{3}, \quad Q_{w} = \tilde{q} \sqrt{2}
\eea
where $\tilde{m}, \tilde{q}$, and $\tilde{r}$ is the mass, charge, and radius of source respectively, and $\tilde{\epsilon}$ is intended to recover the cosmological constituents of the spacetime, and has dimensions of $m^{-2}$. For $\lambda > Q_{w}$, spacetime allows for two horizons; an event, and a cosmological horizon while there is a unique horizon for the extremal BH ($\lambda = Q_{w}$), and $\lambda < Q_{w}$ corresponds to the naked singularity.



\subsubsection{Regular BHs in conformal massive gravity}

The conformal massive gravity is an invariant theory under a conformal transformation of metric tensor as
\beq
g_{\alpha \beta} \rightarrow g^{*}_{\alpha \beta} = \tilde{\Omega} ~ g_{\alpha \beta},
\eeq
where $\tilde{\Omega} = \tilde{\Omega}(x)$ is a nonsingular function of spacetime coordinates. It is noteworthy that the solutions of GR equations are a subset of the solutions of conformal gravity, and also it can describe the DM and DE scenarios. Recently, the regular BH solutions in this gravity has been proposed and the metric components takes the form \citep{Jus-et-al:2020:PhRvD:}
\bea\nonumber
g_{tt} &=& -\left(1 - \frac{2 M r + Q_{c} ~ r^{2-\lambda_{0}}}{\RS}\right),\\
g_{rr} &=& \frac{\RS}{\Delta_{\rm n}}, \quad g_{\theta\theta} = \RS,\nonumber\\
g_{\phi\phi} &=& \sin^{2}\theta \left[r^{2}+a^{2} + a^2 \left(\frac{2 M r + Q_{c} ~ r^{2-\lambda_{0}}}{\RS} \right) \sin^{2} \theta \right], \nonumber\\
g_{t\phi}&=& - 2 a \left( \frac{ M r + Q_{c} ~ r^{2-\lambda_{0}}}{\RS} \right) \sin^{2}\theta,
\label{MetricCoef-Con}
\eea
with
\bea
\Delta_{\rm n} = r^2 - 2 M r + a^2 - Q_{c} ~ r^{2-\lambda_{0}},
\eea
where $Q_{c}$ is the scalar charge, and $\lambda_{0}$ is the hair parameter. The metric is singular at the surface $r = r_{sing}$, where $\RS = 0$.\\

\subsubsection{Regular BHs in Einstein-Yang-Mills theory}

The dynamical interacting system of equations related to the non-abelian gauge theories defined on a curved spacetime is named as Einstein–Yang–Mills theory of gravity which describes the phenomenology of Yang–Mills fields interacting with the gravitational attraction, such as the electro-weak model or the strong nuclear force associated with quantum chromodynamics \citep{Jus-et-al:2021:PhRvD}. The metric co-efficients of the regular, rotating, and magnetic charged BH solution with a Yang-Mills electromagnetic source in the non-minimal Einstein-Yang-Mills theory takes the form
\bea
g_{tt} &=& - \left(1 - \frac{2 r \zeta(r)}{\RS}\right), \quad
g_{rr} = \frac{\RS}{\tilde{\Delta}}, \quad g_{\theta\theta} = \RS,\nonumber\\
g_{\phi\phi} &=& \frac{\sin^{2}\theta}{\RS} \left[(r^2 + a^2)^2 - \tilde{\Delta} a^2 \sin^{2}\theta \right], \nonumber\\
g_{t\phi}&=& -  \frac{2 a  r ~ \zeta(r)}{\RS} \sin^{2}\theta ,
\label{MetricCoef-R}
\eea
with
\bea
     \zeta(r) &=& \frac{1}{2} \left(r \left(1 - Y(r) \right)\right),\\\
     \tilde{\Delta} &=& r^2 + a^2 - \frac{r^6}{(2 \tilde{\lambda} + r^4)} \left(\frac{2 M }{r } - \frac{\tilde{Q}^2  }{ r^2 }\right),\\\
     Y(r) &=& 1+ \frac{r^4}{(2 \tilde{\lambda} + r^4)} \left( \frac{\tilde{Q}^2  }{ r^2 } - \frac{2 M }{r} \right) ,
\eea
where $\tilde{Q}$ is the magnetic charge. This spacetime is free from the singularities and satisfies the energy conditions outside the outer horizon. For $\tilde{\lambda} = 0$, Eq. (\ref{MetricCoef-R}) reduces to the KN BH with a magnetic charge instead of an electric charge, and $\tilde{\lambda} = \tilde{Q} =0$, we obtain the Kerr solution.

\subsubsection{Hairy BHs}

The modified Kerr BH solution so called rotating hairy BHs are surrounded by an axially symmetric “tensor-vacuum” represented by a conserved energy-momentum tensor which could account for one or more fundamental fields (tensor, vector, or scalar fields representing any phenomenologically viable form of matter-energy, such as DM or DE). The energy-momentum tensor satisfies either the dominant energy condition or the strong energy condition in a region outside the event horizon. The metric co-efficients for hairy Kerr BHs can be written as \citep{Con-Ova-Cas:2021:PhRvD:}
\bea
g_{tt} &=& -\left(1 - \frac{2 r m(r)}{\RS}\right), \quad
g_{rr} = \frac{\RS}{\Delta_{\rm h}}, \quad g_{\theta\theta} = \RS,\nonumber\\
g_{\phi\phi} &=& \sin^{2}\theta \left[r^{2}+a^{2} + \frac{2 r a^{2} m(r)}{\RS} \sin^{2} \theta \right], \nonumber\\
g_{t\phi}&=& - \frac{2 a r m(r)}{\RS}\sin^{2}\theta,
\label{MetricCoef-Hair}
\eea
with
\beq
\Delta_{\rm h} = r^2 - 2 r m(r) + a^2,
\eeq
and the mass function $m(r)$ takes the form
\beq
m(r) = M - \frac{r \alpha_{1}}{2} ~ e^{- r/(M-\alpha_{0}/2)}.
\eeq
The parameter $\alpha_{1}$ determines the deviation from the Kerr BH, while the parameter $\alpha_{0}$ measures the increase of entropy caused by the hair and must satisfy the condition $\alpha_{0} \leq 2M $ to ensure asymptotic flatness. For $\alpha_{1} = 0$, hairy BH solution reduces to the classical Kerr BH.

\subsection{Rotating BHs modified by quintessence/matter field}

Recently, with the help of Event Horizon Telescope's observations of BH shadows, it has been proposed that the existence of BHs in the universe is almost universally accepted \citep{EHTSC:2019:ApJ:}. Inspired by this, many physicists have begun to study the interaction between DM (DM) and BHs \citep{Kava-et-el:2020:PhRvD:, Nar-et-al:2020:PhRvD:, Xu-Wang-Tang:2021:JCAP:}. Due to the existence of the supermassive BHs at the centers of galaxies, the strong gravitational potential of the BH concentrates a large amount of DM particles near the BH horizon \citep{Gon-Silk:1999:PhRvL:}. The DM density increases by orders of magnitude due to the BH's gravitational field. Therefore, if DM particles can annihilate into gamma-ray radiation, the intensity of gamma-ray radiation near the BH will increase greatly, which provides a good opportunity for us to detect the DM annihilation signal. A series of DM models have been proposed in literature, some of them we consider here.

\subsubsection{BHs in DM (dirty BHs)}

The rotating BH solution surrounded by a spherical shell of DM can be expressed as \citep{Pan-Rod:2020:ChJPh:}
\bea\nonumber
g_{tt} &=& -\left(1 - \frac{2 r m(r)}{\RS}\right), \quad
g_{rr} = \frac{\RS}{\Delta_{\rm d}}, \quad g_{\theta\theta} = \RS,\nonumber\\
g_{\phi\phi} &=& \sin^{2}\theta \left[r^{2} + a^{2} + \frac{2 r a^{2} m(r)}{\RS} \sin^{2} \theta \right], \nonumber\\
g_{t\phi}&=& - \frac{2 a r m(r)}{\RS}\sin^{2}\theta,
\label{MetricCoef-DM}
\eea
with
\bea
\Delta_{\rm d} &=& r^2 - 2 r (M + \Delta M) H(r) + a^2, \nonumber\\
H(r) &=& \left(3-\frac{2 \left(r - r_s \right)}{\Delta r_s} \right) \left(\frac{r - r_s}{\Delta r_s} \right)^2,
\eea
where the piecewise continuous mass function $m(r)$, written in the form
\[ m(r)= \begin{cases}
      M, & r < r_{s}; \\
      M + \Delta M ~ H(r), & r_{s} \leq r \leq r_{s} + \Delta r_{s}; \\
      M + \Delta M, & r > r_{s} + \Delta r_{s}.
   \end{cases}
\]
Here, $\Delta M < 0$ and $\Delta M > 0$ indicates the positive and negative energy density of matter, while $r_{s}$, and $\Delta r_{s}$ represent the inner radius, and thickness of the spherical shell of DM, respectively, for details see \citep{Konoplya:2019:PhLB:}.

\subsubsection{BHs in perfect fluid DM}

The non-zero metric co-efficients of rotating BH in perfect fluid DM can be written as \citep{Hou-Xu-Wan:2018:JCAP}
\bea \non
g_{tt} &=& -\left(1 - \frac{2Mr - f(r)}{\RS}\right),\non\\
g_{rr} &=& \frac{\RS}{\Delta_{\rm D}}, \quad g_{\theta\theta} = \RS,\nonumber\\
g_{\phi\phi} &=&  \left[r^{2} + a^{2} + \frac{a^{2}\left(2 M r - f(r) \right )\sin^{2} \theta}{\RS} \right]\sin^{2}\theta, \nonumber\\
g_{t\phi}&=& - \frac{a \left(2 M r - f(r) \right)}{\RS}\sin^{2}\theta,
\label{MetricCoef-DMfluid}
\eea
where
\bea
\Delta_{\rm D} &=& r^2 - 2 M r + a^{2} + f(r),\\
f(r) &=& r k ~ \ln\left(\frac{r}{|k|}\right).
\eea
The parameter $k$ determines the perfect fluid DM intensity, and in the absence of perfect fluid DM, one can recover the Kerr metric.

\subsubsection{BHs in cold DM halo}

The strong gravity of a supermassive BH in the center of a galaxy could enhance the DM density significantly, producing a phenomenon known as ``spike" \citep{Gon-Silk:1999:PhRvL:}. But for the Navarro-Frenk-White density profile, a ``cusp" problem occurs \citep{Blok:2010:AdAst:2010:} - a contradiction to the observations which show rather a flat density profile. However, for other DM models, i.e., scalar field dark DM, modified newtonian dynamics DM and warm DM, ``cusp" is not produced in small scale. Motivated by these problems, Xu-et-al \citep{Xu-et-el:2018:JCAP:} proposed the rotating BHs surrounded by DM halos, solution for the cold DM halos read
\bea\non
    g_{tt} &=& - \left(1 - \frac{r^{2} + 2 M r - r^{2} \mathbb{X}}{\RS}\right),\quad
    g_{rr} = \frac{\RS}{\Delta_{\rm c}}, \nonumber\\
     g_{\theta\theta} &=& \RS, \quad
    g_{\phi\phi} =  \left[(r^2 + a^2)^2 -  a^2 \sin^{2}\theta \Delta_{\rm c} \right] \frac{\sin^{2}\theta}{\RS} , \nonumber\\
    g_{t\phi}&=& -  \frac{2 a \sin^{2}\theta}{\RS} \left(r^2 + 2 M r - r^{2} \mathbb{X}  \right),
\label{CDM-Halo}
\eea
with
\beq
\Delta_{\rm c} = r^2 \mathbb{X} + a^2 -2 M r, \quad \mathbb{X} = \left(1 +  \frac{r}{\rm R_c}\right)^{\frac{-8 \pi R_{\rm c}^{3} \rho_{c} }{r}}.
\eeq
Here, $\rho_{c}$ denotes the density of the Universe at the moment when the halo collapsed and $R_{\rm c}$ is the characteristic radius. For vanishing cold DM, Eq. (\ref{CDM-Halo}) reduces to the Kerr BH with DM.

\subsubsection{BHs in scalar field DM halo}

The rotating BHs surrounded by scalar field DM halo can be described by \citep{Xu-et-el:2018:JCAP:}
\bea\non
    g_{tt} &=& - \left(1 - \frac{r^{2} + 2 M r - r^{2} \mathbb{Y}}{\RS}\right),\quad
    g_{rr} = \frac{\RS}{\Delta_{\rm s}}, \nonumber\\
     g_{\theta\theta} &=& \RS, \quad
    g_{\phi\phi} =  \left[(r^2 + a^2)^2 -  a^2 \sin^{2}\theta \Delta_{\rm s} \right] \frac{\sin^{2}\theta}{\RS} , \nonumber\\
    g_{t\phi}&=& -  \frac{2 a \sin^{2}\theta}{\RS} \left(r^2 + 2 M r - r^{2} \mathbb{Y}  \right),
\label{SFDM-Halo}
\eea
with
\bea
\Delta_{\rm s} &=& r^2 ~ \mathbb{Y} + a^2 -2 M r,\\\
\mathbb{Y} &=& \exp \left[ - \frac{8 \rho_{s} R_{s}^{2}}{\pi} \frac{\sin(\pi r/R_{\rm s})}{\pi r/ R_{\rm s}}\right],
\eea
where, $\rho_{s}$ denotes the density of the Universe at the moment when the halo collapsed and $R_{\rm s}$ is the characteristic radius.

\subsubsection{Hayward BHs in perfect fluid DM}

The solutions of the Einstein equations coupled to a nonlinear electromagnetic field in the presence of PFDM, representing the rotating and non-linear magnetic charged BHs surrounded by PFDM are given by \citep{Ma-etal:2021:MPLA:}
\bea \non
g_{tt} &=& -\left(1 - \frac{r^2 - z(r) r^2}{\RS}\right),\non\\
g_{rr} &=& \frac{\RS}{\Delta_{\rm h}}, \quad g_{\theta\theta} = \RS,\nonumber\\
g_{\phi\phi} &=&  \left[(r^{2} + a^{2})^2 - a^2 \Delta_{\rm h} \sin^{2}\theta \right] \frac{\sin^{2}\theta}{\RS}, \nonumber\\
g_{t\phi}&=& - \frac{a \left(r^2 - z(r) r^2 \right)}{\RS}\sin^{2}\theta,
\label{MetricCoef-Hayward-DMfluid}
\eea
where
\bea
\Delta_{\rm h} &=& r^2 z(r)+ a^{2},\\
z(r) &=& 1 - \frac{2 M r^2}{r^3 + Q_{\rm h}^3} + \frac{\tilde{k}}{r} \ln \left(\frac{r}{|\tilde{k}|}\right),
\eea
where $Q_{\rm h}$ is the magnetic charge of BH, and $\tilde{k}$ denotes the intensity of PFDM. The value of $\tilde{k}$ can be both positive and negative. In the absence of PFDM ($\tilde{k} = 0$), one can obtain the rotating, non-linear magnetic charged BH, and for $\tilde{k} = Q_{\rm h} = 0$, Eq. (\ref{MetricCoef-Hayward-DMfluid}) reduces to the Kerr BH.

\subsubsection{BHs in DM spike}

If the galactic center contains DM, then the existance of a supermassive BH in the galactic center would produce a cusp in the distribution of DM, known as DM spike \citep{Gon-Silk:1999:PhRvL:}. In order to study the effects of DM spike on BHs, Nampalliwar et al. \citep{Sou-etal:2021:ApJ:}, proposed the rotating BH solutions immersed in DM spike, given by
\bea \non
g_{tt} &=& - \frac{\mathbb{H}(r)}{\RS_{\rm sp}} \left (\frac{a^2 \sin ^2 \theta - \Delta_{\rm sp}}{\RS_{\rm sp}} \right),\non\\
g_{rr} &=& \frac{\mathbb{H}(r)}{\Delta_{\rm sp}}, \quad g_{\theta\theta} = \mathbb{H}(r),\nonumber\\
g_{\phi\phi} &=& \frac{\mathbb{H}(r)}{\RS_{\rm sp}} \left( \frac{ \left(a^2 + \mathbb{K}(r) \right)^2 - a^2 \Delta_{\rm sp} \sin ^2\theta }{\RS_{\rm sp}} \right) \sin^2 \theta , \nonumber\\
g_{t\phi}&=& - \frac{\mathbb{H}(r)}{\RS_{\rm sp}} \left( \frac{a \left(a^2 + \mathbb{K}(r) \right) - a \Delta_{\rm sp}}{\RS_{\rm sp}} \right)\sin^2 \theta,\label{MetricCoef-DM-spike}
\eea
with
\bea\non
\mathbb{F}(r) &=& 1- \frac{2 M}{r} - \exp\left[-\frac{8 \pi  R_{\rm b}^2 ~ \rho_d \left(\frac{R_{\rm sp}}{R_{\rm b}}\right){}^{\gamma_{\rm s}}}{\left(\gamma_{\rm s} - 2 \right)} \right] \\\ &+& \exp[\frac{- 8 \pi  \rho_{\rm d} \left(R_{\rm b}^3 \left(\gamma_{\rm s} - 2\right) \left(\frac{R_{\rm sp}}{R_{\rm b}}\right){}^{\gamma_{\rm s}} - r^3 \left(\frac{R_{\rm sp}}{r}\right){}^{\gamma_{\rm s}}\right)}{r \left(\gamma_{\rm s} - 3 \right) \left(\gamma_{\rm s} - 2 \right)}] ,\\\non
\mathbb{G}(r) &=&  1 - \frac{2 M}{r} + \frac{\left(8 \pi ~ r^2 \rho_{\rm d} \right) \left(\frac{R_{\rm sp}}{r}\right)^{\gamma_{\rm s}}}{(\gamma_{\rm s} - 3)} \\\ &-& \frac{\left(8 \pi  R_{\rm b}^3 \rho_{\rm d} \right) \left(\frac{R_{\rm sp}}{R_{\rm b}} \right){}^{\gamma_{\rm s}}}{r \left(\gamma_{\rm s} - 3\right)},\\\
\mathbb{H}(r) &=& \sqrt{\mathbb{G}(r)/ \mathbb{F}(r)}~ r^2 + a^2 \cos^2 \theta,\\\
\mathbb{Z}(r) &=& r^2, \quad \mathbb{K}(r) = \sqrt{\mathbb{G}(r)/ \mathbb{F}(r)} ~ \mathbb{Z}(r),\\\
\Delta_{\rm sp} &=& a^2 + \mathbb{Z}(r) \mathbb{G}(r),\\\
\RS_{\rm sp} &=& \mathbb{K}(r) + a^2 \cos^2 \theta,
\eea
where $\gamma_{\rm s} = (9 - 2 \gamma)/(4 - \gamma)$, $\gamma$ is the power-law index, $\rho_{\rm d}$ denoted the density of DM spike, $R_{\rm sp}$ shows the radius of DM spike, and $R_{\rm b}$ is the inner edge of the DM spike. For details, see \citep{Sou-etal:2021:ApJ:}.

\subsubsection{Deformed BHs in DM spike}

Recently, Xu et al. developed the deformed BH solutions immersed in DM spike, given by \citep{Xu-etal:2021:JCAP:}
\bea \non
g_{tt} &=& - \left(\frac{r^2 G(r) + a^2 \cos^{2}\theta }{\RS} \right),\non\\
g_{rr} &=& \frac{\RS}{\Delta_{\rm sd}}, \quad g_{\theta\theta} = \RS,\nonumber\\
g_{\phi\phi} &=& \RS \left[1 + a^2 \sin^{2}\theta ( \frac{2 r^2 + a^2 \cos^{2}\theta - r^2 G(r) }{\RS}) \right]\sin^{2}\theta, \nonumber\\
g_{t\phi}&=& - \frac{a \sin^{2}\theta (1 - G(r) ) r^2}{\RS},\label{MetricCoef-DM-spike}
\eea
where
\bea\non
\Delta_{\rm sd} &=&  a^{2} + r^2 G(r),\\\non
 &=& a^2 - 2 M r + r^{\frac{48 \pi  \rho_{\rm R} (R_{s\rm p})^{\tilde{\alpha }} \left(k_{0} R_{\rm s} \right){}^{3 - \tilde{\alpha }}}{\tilde{\alpha } \left(\tilde{\alpha } - 1 \right) \left(\tilde{\alpha } - 2 \right) \left(\tilde{\alpha } - 3\right)} + 2} \\\non & \times &  \exp[ - \frac{r^{ - \tilde{\alpha }} \left(8 \pi  \rho_{\rm R} \left(k_{0} R_{\rm s} \right){}^3 R_{\rm sp}^{\tilde{\alpha }} \right)}{\tilde{\alpha }^2} \\\non &+& \frac{r^{1 - \tilde{\alpha }} \left(24 \pi  \rho_{\rm R} \left(k_{0} R_{\rm s}\right){}^2 R_{\rm sp}^{\tilde{\alpha }}\right)}{\left(\tilde{\alpha } - 1 \right)^2} \\\non &-& \frac{r^{2 - \tilde{\alpha }} \left(24 \pi  \rho_{\rm R} \left(k_{0} R_{\rm s} \right) R_{\rm sp}^{\tilde{\alpha }} \right)}{\left(\tilde{\alpha } - 2\right)^2} \\\ &+& \frac{r^{3 - \tilde{\alpha }} \left(8 \pi  \rho_{\rm R} R_{\rm sp}^{\tilde{\alpha }} \right)}{\left(\tilde{\alpha } - 3\right)^2}].
\eea
Here, $\rho_{\rm R}$ denotes the normalization of the DM density, $\tilde{\alpha }$ represents the power-law index, $k_{0}$ is the DM zero point parameter, $R_{\rm sp}$ shows the radius of the DM spike, and $R_{\rm s}$ is the Schwarzschild radius of BH. In the absence of DM spike ($\rho_{\rm R} = 0$), this BH reduces to the usual Kerr BH.

\subsubsection{BHs in quintessence}

The quintessence is dynamical and inhomogeneous scalar field having negative pressure, fully characterized by the equation $\rho = w p$, where $\rho$ and $p$ indicates the energy density and pressure, respectively. The non-zero metric co-efficients of rotating BH solution surrounded by quintessence can be expressed as \citep{Xu-Wang:2017:PhRvD:,Tos-Stu-Ahm:2017:EPJP:,Ift-Sha:2019:EPJC}
\bea
g_{tt} &=& -\left(1 - \frac{2 M r + \tilde{c} ~ r^{1-3 \tilde{\omega}}}{\RS}\right), \quad
g_{rr} = \frac{\RS}{\Delta_{\rm q}}, \quad g_{\theta \theta} = \RS,\nonumber\\
g_{\phi\phi} &=& \sin^{2}\theta \left[r^{2} + a^{2} + a^{2} \left( \frac{2 M r + \tilde{c} ~ r^{1-3 \tilde{\omega}}}{\RS} \right) \sin^{2} \theta \right], \nonumber\\
g_{t\phi}&=& - a^2 \left( \frac{2 M r + \tilde{c}~ r^{1 - 3 \tilde{\omega}}}{\RS} \right)\sin^{2}\theta,
\label{MetricCoef-Bar}
\eea
with
\beq
\Delta_{\rm q} = r^2 - 2 M r + a^2 - \tilde{c} ~ r^{1 - 3 \tilde{\omega}},
\eeq
where $\tilde{c}$ is the quintessential field parameter. There are three cases according to the value of state parameter $\tilde{\omega}$, i.e., $\tilde{\omega} < -1$, $\tilde{\omega} = 1$, and $ -1 < \tilde{\omega} < -1/3 $ , corresponds to the phantom energy, the cosmological constant and the quintessence, respectively \citep{She-etal:2020:PhRvD:}.

\section{Orbital Period} 

The equations of motion for test particles in an alternative theories of gravity need not be geodesic. However, in the test-particle limit, equations of motion can be approximated as geodesics for a wide class of alternative theories, neglecting the spin of the small body \citep{Vig-Yun-Ste:2011:PhRvD}. Here, we restrict our attention to theories where the modified equations of motion remain geodesic.

The orbital frequency describes the motion of test particles in the azimuthal direction, observed at radial infinity, defined by $\Omega_{\phi} = \dot{\phi} / \dot{t}$ (where dot represents the derivative with respect to proper time $\tau$), and can be found with the help of the geodesic equation
\beq
\frac{d^2 x^\mu}{d \tau} = \Gamma^{\mu}_{\alpha \beta} \frac{d x^{\alpha}}{d \tau} \frac{d x^{\beta}}{d \tau},
\eeq
where $\Gamma^{\mu}_{\alpha \beta}$ are the Christoffel symbols which can be written in the form
\beq
\frac{d }{d \tau} \left( g_{\beta \eta} \frac{d x^{\eta}}{d \tau} \right) = \frac{1}{2} \frac{\partial g_{\alpha \delta}}{\partial x^{\beta} } \frac{d x^{\alpha}}{d \tau} \frac{d x^{\delta}}{d \tau}.
\eeq
Due to the reflection and axi-symmetric properties of the spacetime, for the existence of equatorial circular orbits, we have
\beq
\frac{d r}{d \tau} = \frac{d \theta}{d \tau} = \frac{d^2 r}{d \tau^2 } = 0.
\eeq
Consequently, the equation that describes the particle motion in radial direction reduces to the following relation
\beq
g_{tt,r}\dot{t}^2 + 2 g_{t \phi, r}\dot{t} \dot{\phi} + g_{\phi \phi, r}\dot{\phi}^2 = 0,
\eeq
Thus, the orbital frequency can be written the form
\beq
\Omega_{\phi} = \frac{-g_{t\phi,r} \pm \sqrt{(g_{t\phi,r})^2 - g_{tt,r} ~ g_{\phi\phi,r}}}{g_{\phi\phi,r}},\label{Freq-1}
\eeq
where the upper and lower signs refer to the prograde, and retrograde orbits respectively. It is clear from Eq. (\ref{Freq-1}), the orbital frequency $\Omega_{\phi}$ in independent of the metric co-efficients $g_{rr}$ and $g_{\theta \theta}$, while the partial derivatives of $g_{tt}, g_{\phi \phi}$, and $g_{tr}$ with respect to the radial distance $r$ are involved. The orbital frequency for those BHs having metric coefficients where some extra parameter is not multiplied with the radial distance $r$, is the same as for Kerr BH. We consider only those BHs which have different orbital frequency from that of Kerr BH.
In the flare observations, the period of the orbital motion $P$ is established, that is related to the orbital (Keplerian) frequency $\Omega_{\phi}$ by relation
\beq
P =  \left( \frac{2\pi}{60} \right) \left( \frac{G M}{c^3} \right) \frac{1}{\Omega_\phi}, \label{period}
\eeq
where period $P$ is in minutes. This formula, hot-spot period as function of radius $P(r)$, will be used for fitting observed flare period radius data in the following section.

Using the normalization condition $p^\alpha p_\alpha = -\mu^2$, evaluated in an equatorial plane, the effective potential $V_{\rm eff} (r)$ can be expressed in the form
\beq
V_{\rm eff}(r) = -\frac{1}{2} \left(\mu^2 + g^{tt} E^2 + g^{\phi \phi} L^2 \right) + g^{t \phi} E L,
\eeq
where $E = -p_t$, and $L = p_{\phi}$ are interpreted as energy and axial angular momentum of a particle associated with Killing vector fields $\xi^{\mu}_{(t)}$, and $\xi^{\mu}_{(\phi)}$, respectively. Effective potential is very important since it enables us to demonstrate the general properties of test particle dynamics, avoiding the necessity to solve the equations of motion. The circular equatorial orbits are governed by the condition \citep{Kol-Stu-Tur:2015:CQGra,Kol-Tur-Stu:2017:EPJC}
\beq
V_{\rm eff}(r) = 0, \quad \frac{\d V_{\rm eff} (r)}{\d r} = 0.\label{Veff-1}
\eeq
The energy $E$ and angular momentum $L$ of circular orbits can be found by solving the Eq. (\ref{Veff-1}), and the orbital frequency in terms of constants of motion can be written as
\beq
\Omega_{\phi} = \frac{p_{\phi}}{p_{t}} = - \frac{g_{tt} L + g_{t \phi} E}{g_{t \phi} L + g_{\phi \phi } E}.\label{Freq-2}
\eeq
Combining Eqs. (\ref{Freq-1}), (\ref{Freq-2}), along with the condition $V_{\rm eff} (r) = 0$, one can find the energy and angular momentum in terms of orbital frequency as
\bea
E &=&   \frac{- g_{tt} - g_{t \phi} \Omega_{\phi}}{\sqrt{ - g_{tt} - 2 g_{t \phi} \Omega_{\phi} - g_{\phi \phi} \Omega_{\phi}^{2} }},\\\
L &=& \pm \frac{g_{t \phi} + g_{\phi \phi} \Omega_{\phi}}{\sqrt{ - g_{tt} - 2 g_{t \phi} \Omega_{\phi} - g_{\phi \phi} \Omega_{\phi}^{2}}},
\eea
where the upper and lower signs correspond to the prograde and retrograde orbits, respectively. The smallest stable equatorial circular orbits so called innermost stable circular orbits (ISCO) are governed by the Eq. (\ref{Veff-1}) along with the condition $d^{2}V_{\rm eff} (r)/d r^{2} = 0$. The position of ISCO is one of the parameters that are very sensitive to the value of the BH spin. The location of ISCO for Schwarzschild BH (non-rotating) is situated at $r=6$ from singularity. For rotating BHs, the ISCO of counter-rotating orbits move outwards the BH, while the position of ISCO for co-rotating orbits shift towards the BH.

\begin{figure*}
\begin{center}
\includegraphics[width=0.45\hsize]{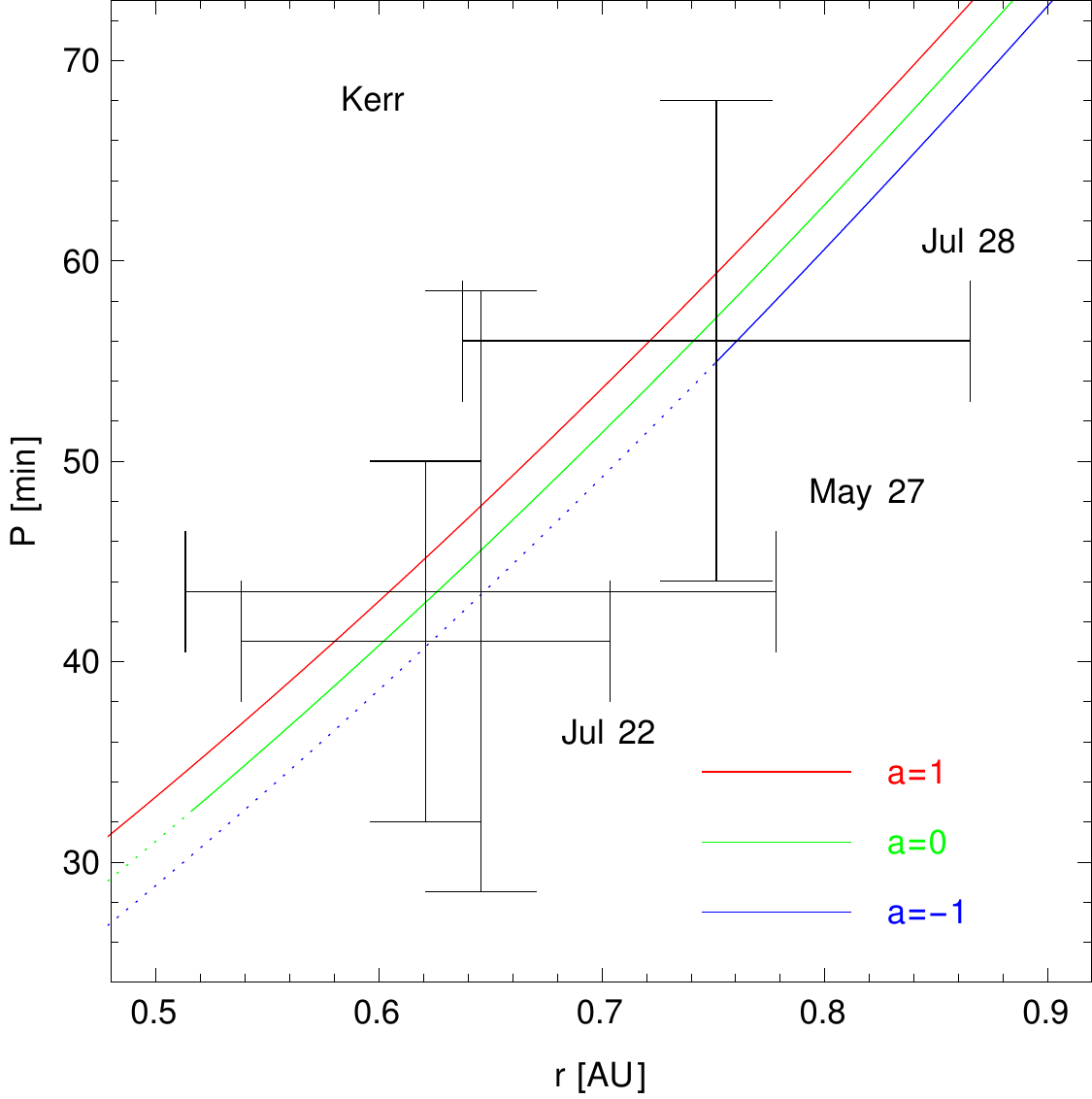}\qquad
\includegraphics[width=0.45\hsize]{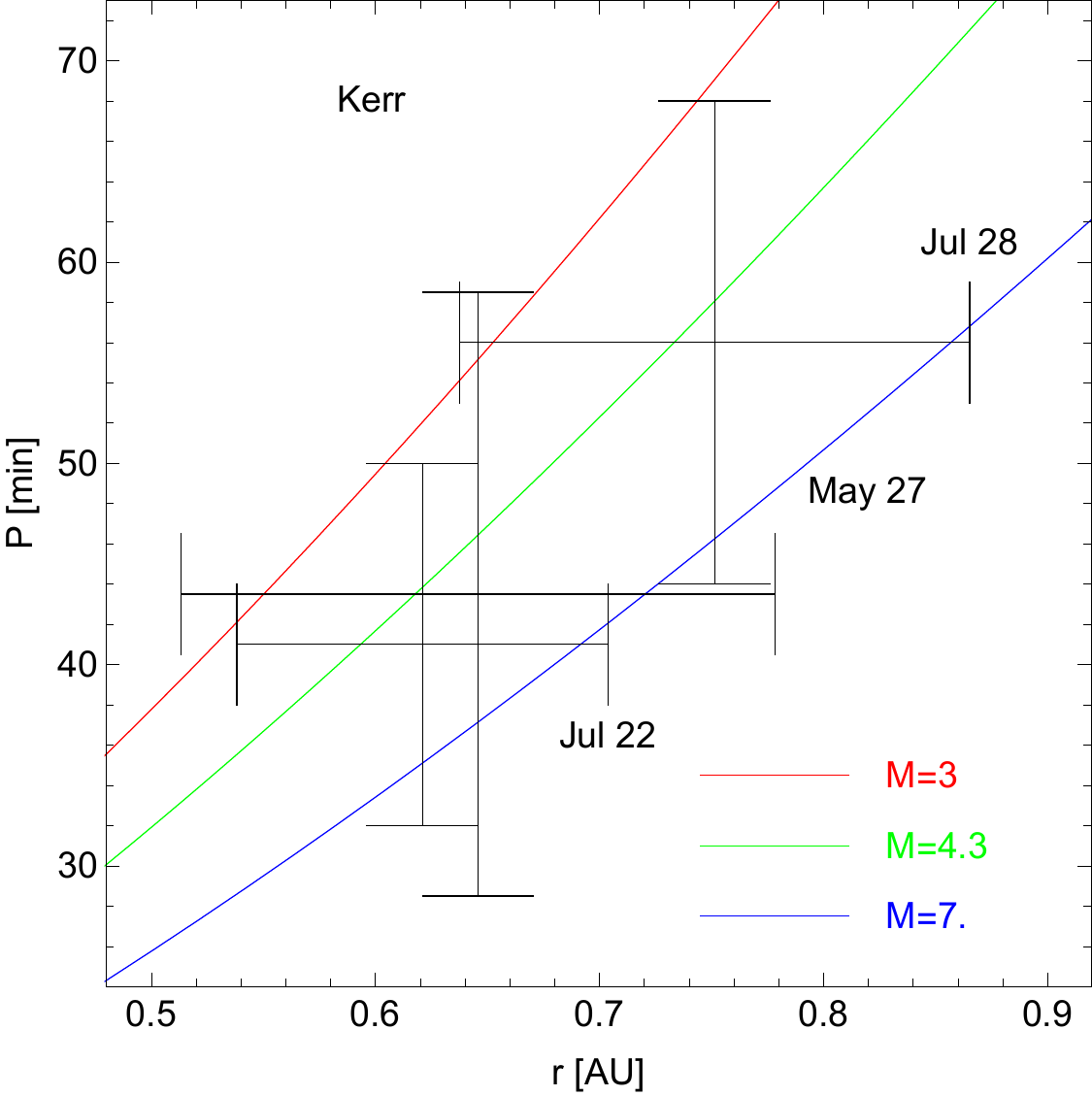}
\end{center}
\caption{Left: The relation between orbital period–radius, and position of three flares proposed by GRAVITY on Jully~22, May~27, and July~28, 2018 fitted with circular orbits of a neutral hot-spot in the background of Kerr BH with mass $M=4.3\times 10^{6}\Msun$. The green curve is plotted for vanishing spin ($a=0$) that corresponds to the Schwarzschild BH, while red and blue curves correspond to the co-rotating and counter-rotating orbits of hot-spot orbiting Kerr BH with spin $a=\pm1$, fitting the observed positions and periods of the flares. The starting points of solid curves represent the ISCO positions, while dotted, and solid parts of the curves are plotted for below and above the ISCO position, respectively. Right: Orbits of neutral hot-spot moving around Kerr BH with three different BH masses $M=(3,4.3,7)\times10^{6}\Msun$, but having same spin $a=0.4$, fitting the observed positions and periods of the flares.
\label{Kerr-Fig}
}
\end{figure*}

\begin{figure*}
\centering
\includegraphics[width=0.3\hsize]{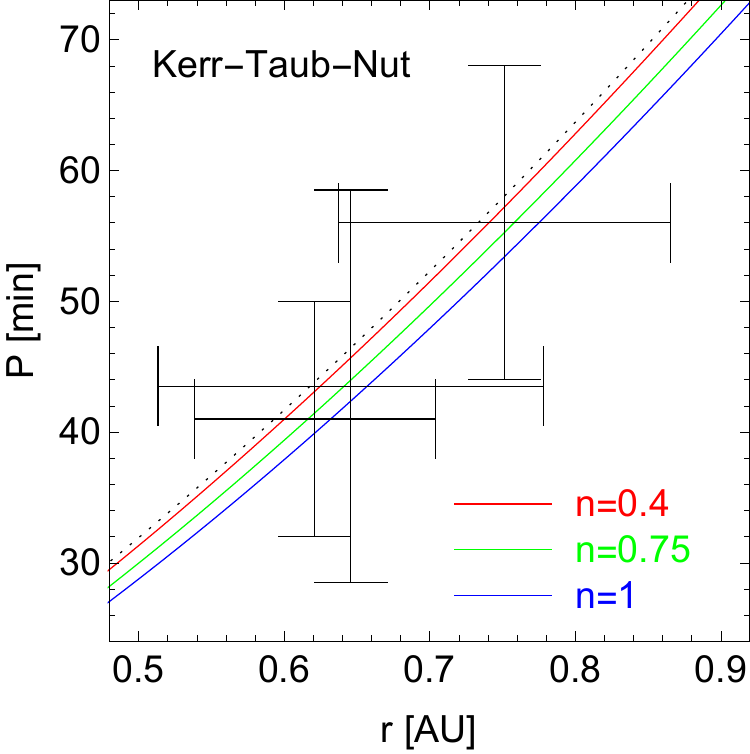}\qquad
\includegraphics[width=0.3\hsize]{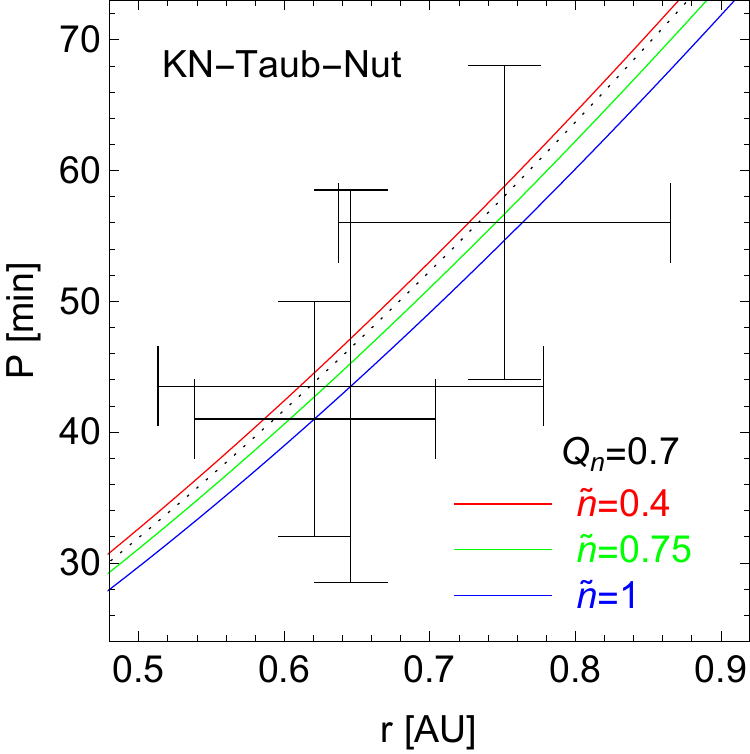}\qquad
\includegraphics[width=0.3\hsize]{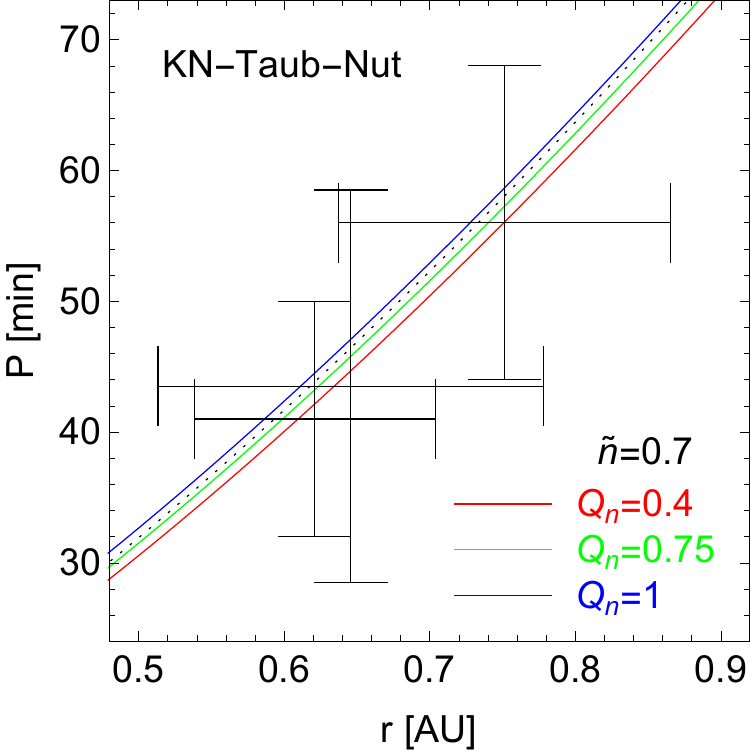}
\includegraphics[width=0.3\hsize]{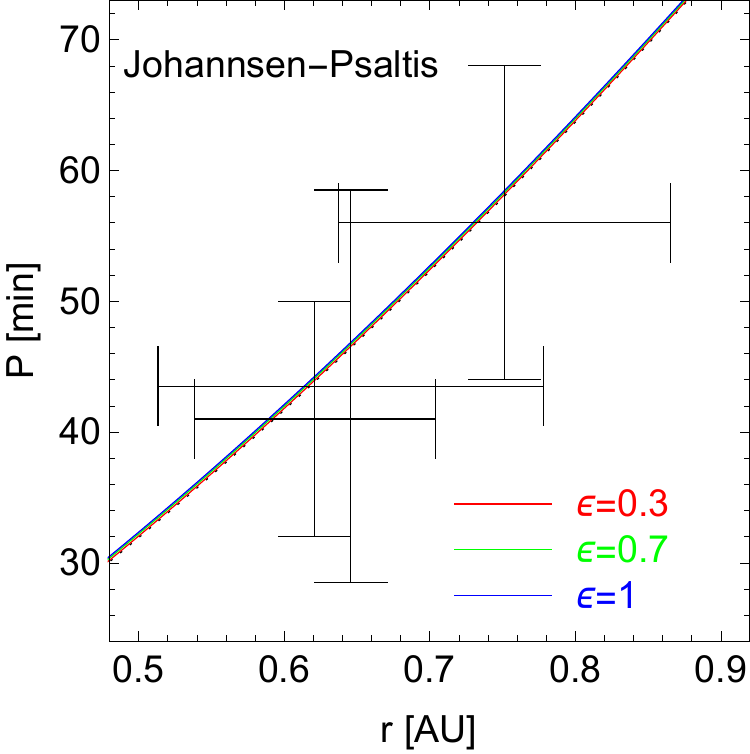}\qquad
\includegraphics[width=0.3\hsize]{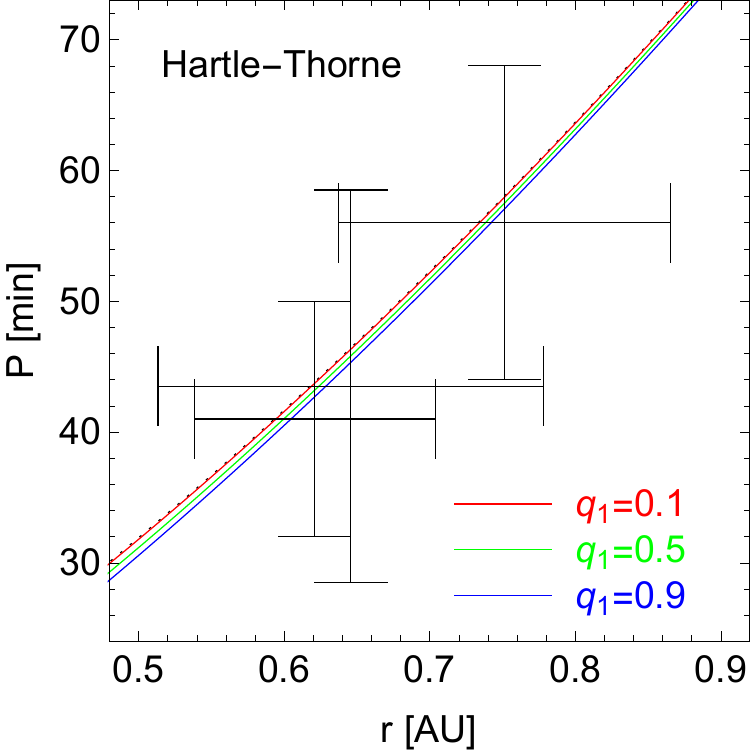}\qquad
\includegraphics[width=0.3\hsize]{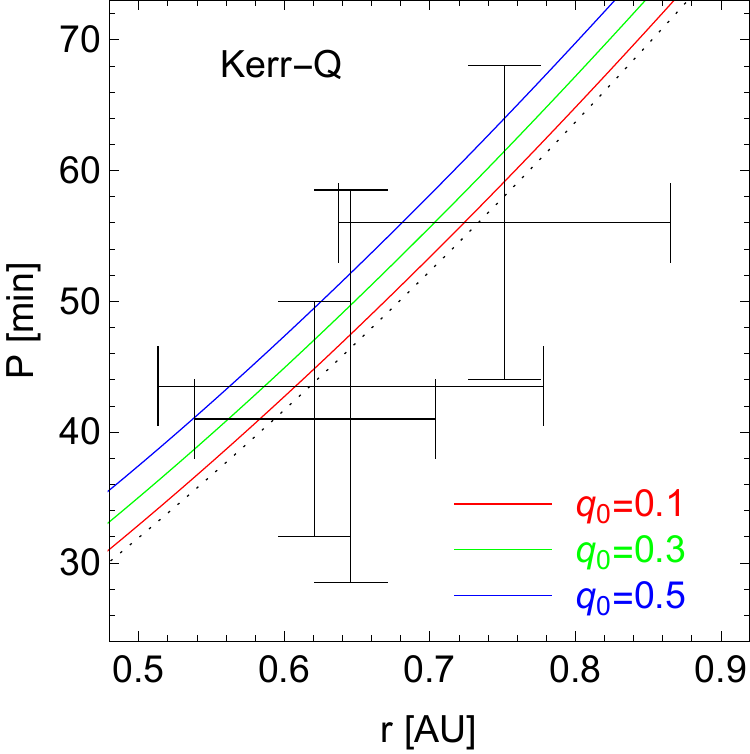}
\includegraphics[width=0.3\hsize]{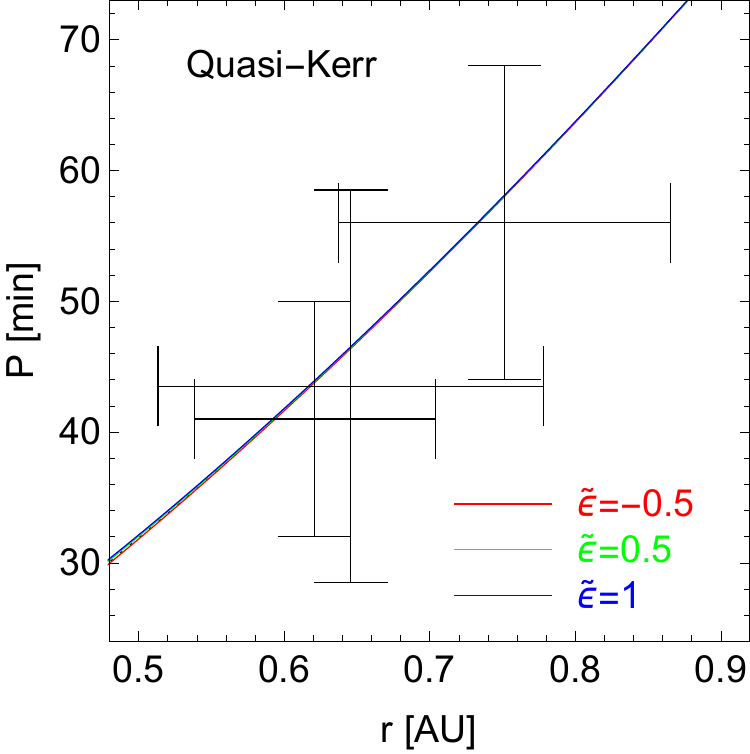}\qquad
\includegraphics[width=0.3\hsize]{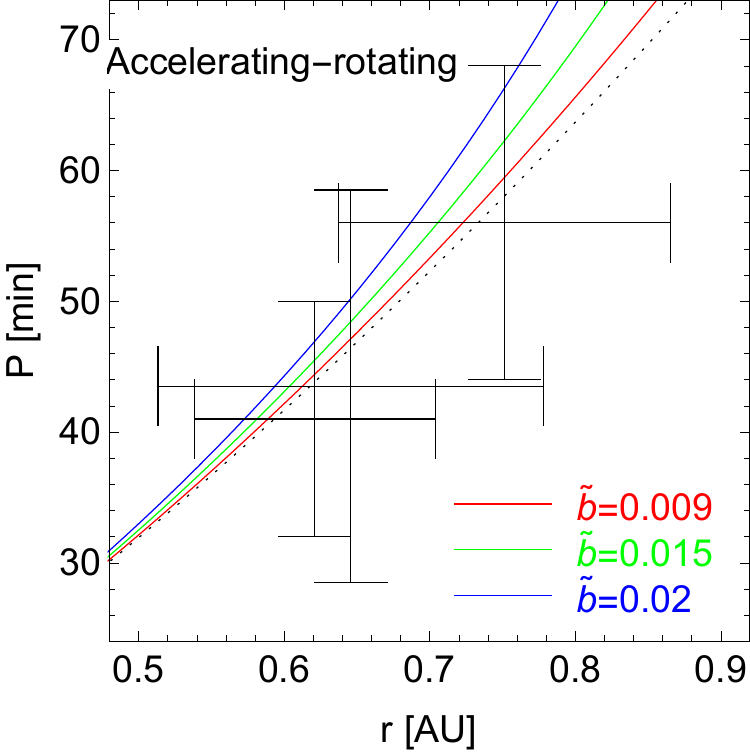}\qquad
\includegraphics[width=0.3\hsize]{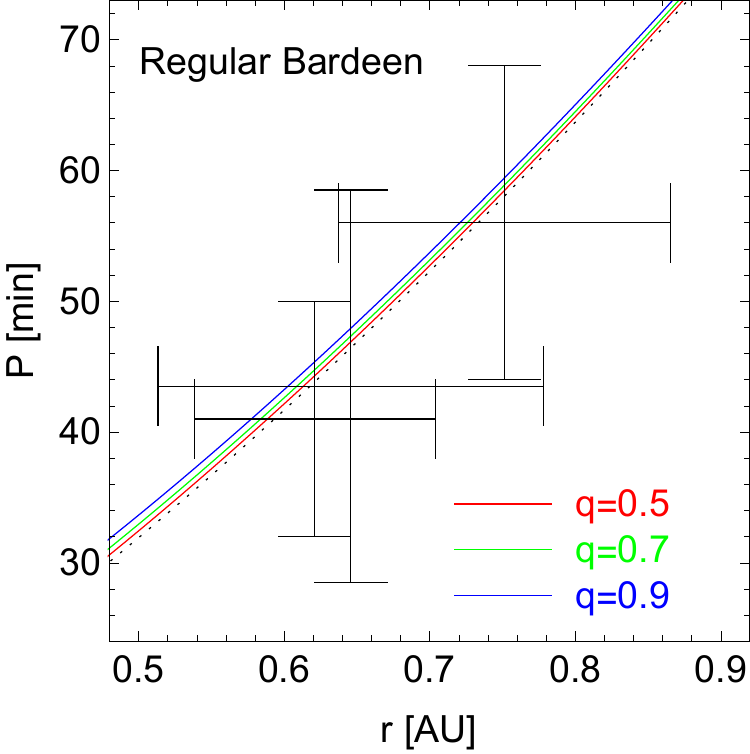}
\caption{The relation between orbital period–radius, and position of three flares proposed by GRAVITY on July~22, May~27, and July~28, 2018 fitted with circular orbits of a neutral hot-spot for many different BHs both in GR as well as alternative theories of gravity. We choose the spin parameter $a=0.4$, and mass $M=4.3\times10^{6}\Msun$ for all BHs. Each plot represents three different orbits fitting the observed positions and periods of three flares. The black dotted curves indicate the Kerr limit. The red curves move towards the blue curves with the increase of the corresponding parameter. The text with each plot represents the designation of BH, and the values of parameters of corresponding BHs, used for the plots are also shown. We see that the circular equatorial orbits for most of the BHs are situated above the center of error bars of all three flares.
\label{fig_all}}
\end{figure*}
\renewcommand{\thefigure}{2}
\begin{figure*}
\centering
\includegraphics[width=0.3\hsize]{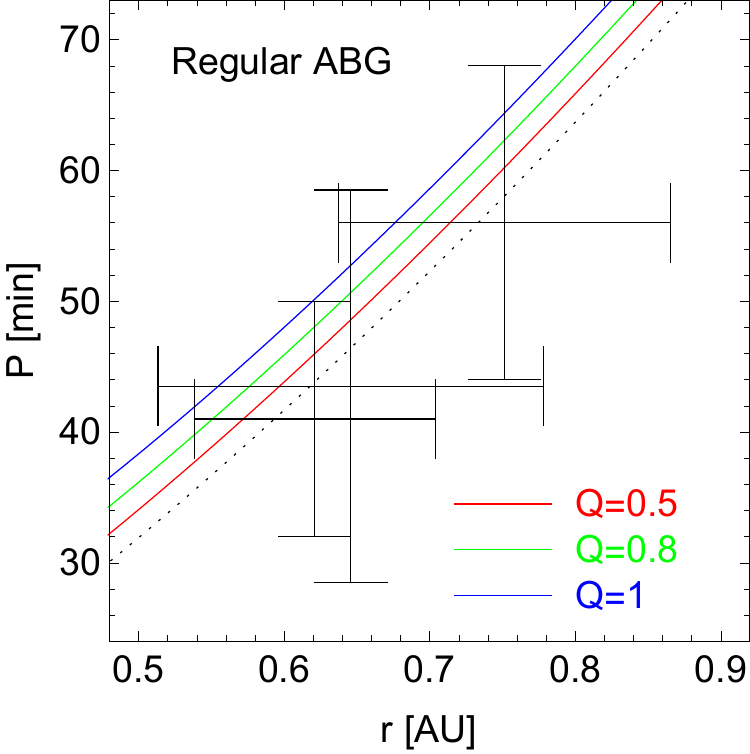}\qquad
\includegraphics[width=0.3\hsize]{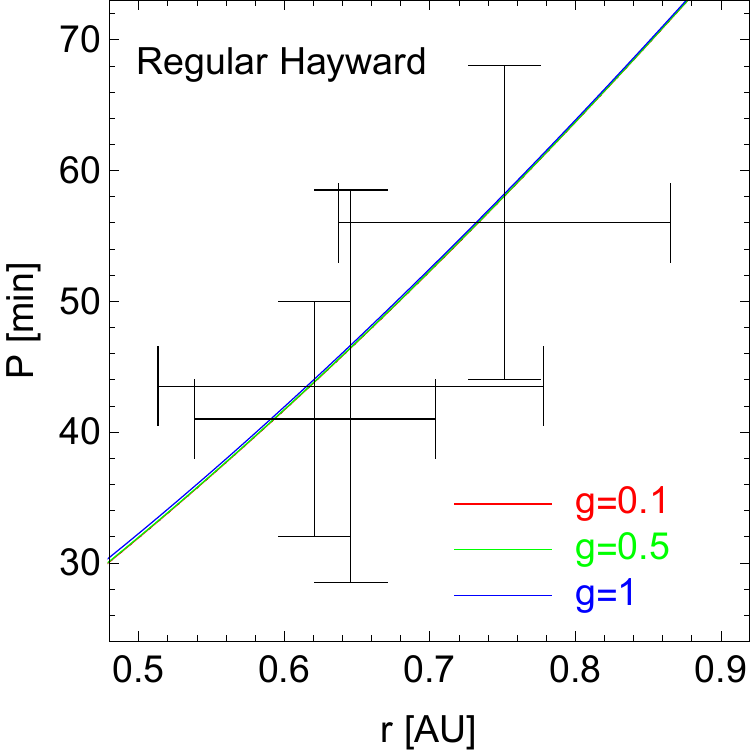}\qquad
\includegraphics[width=0.3\hsize]{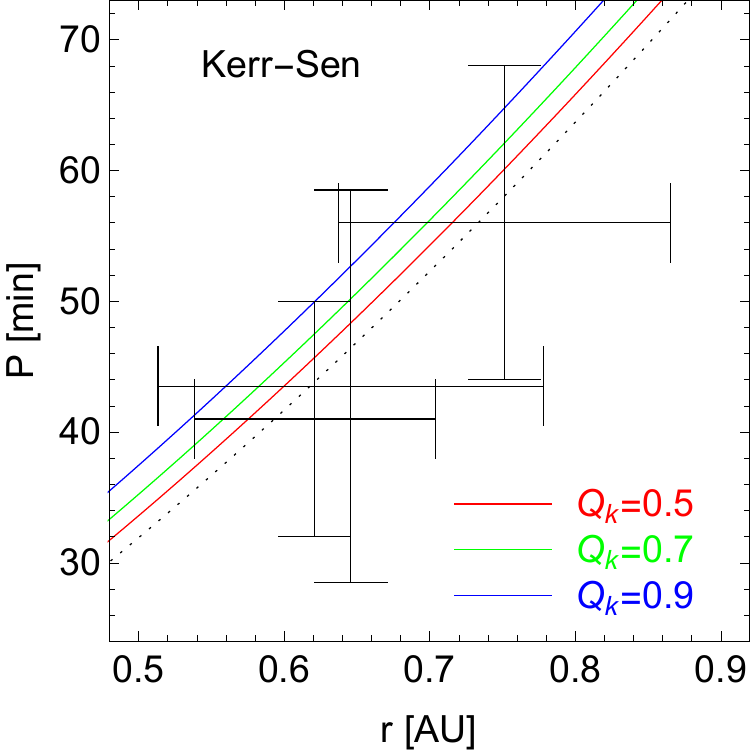}
\includegraphics[width=0.3\hsize]{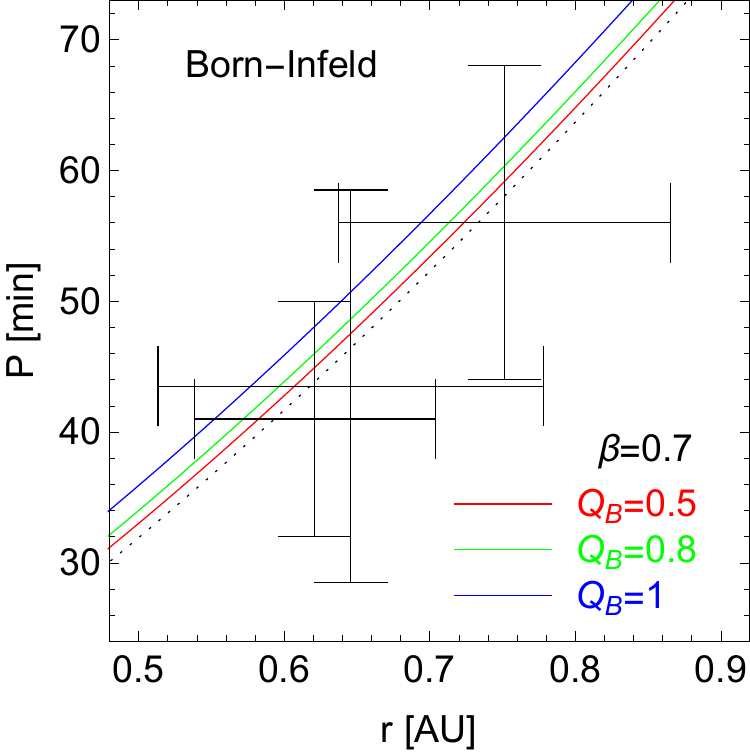}\qquad
\includegraphics[width=0.3\hsize]{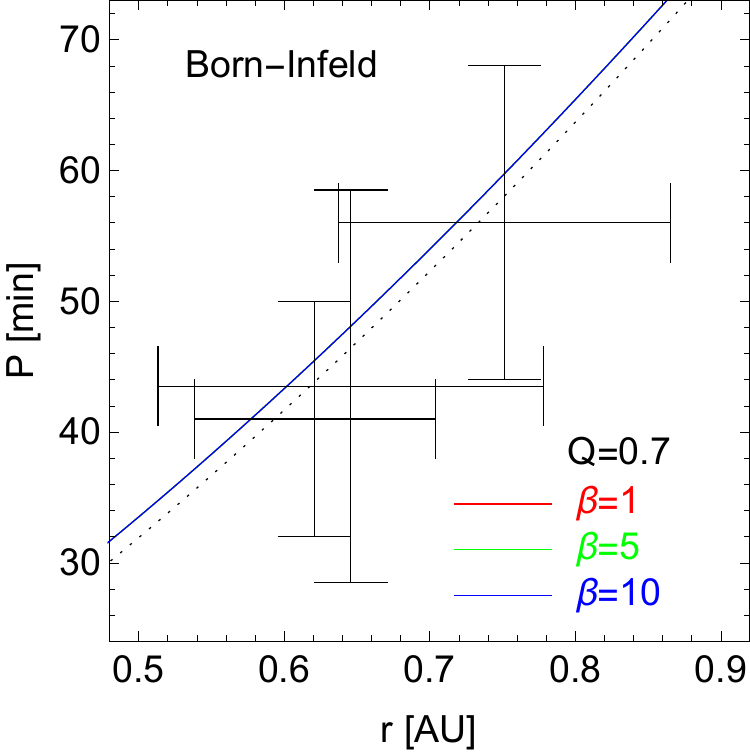}\qquad
\includegraphics[width=0.3\hsize]{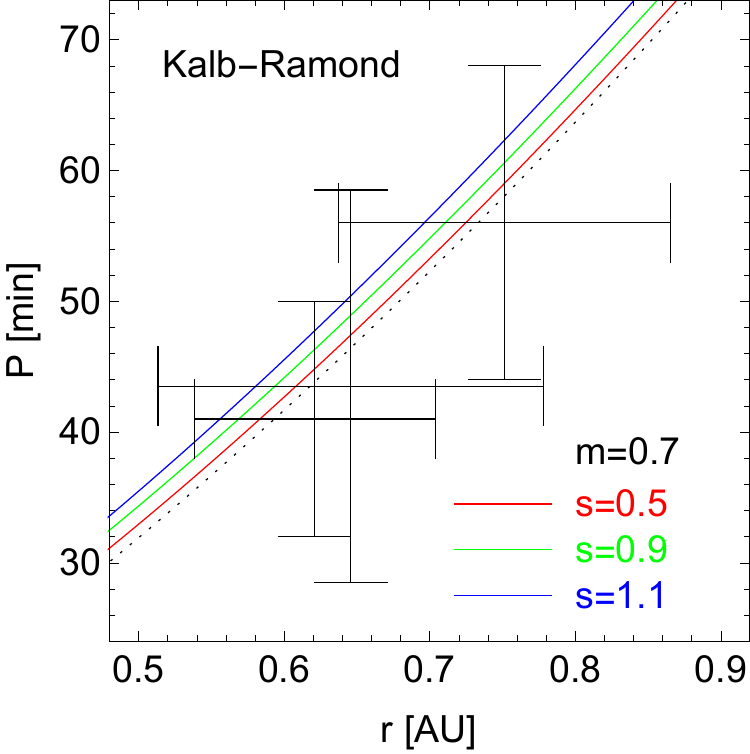}
\includegraphics[width=0.3\hsize]{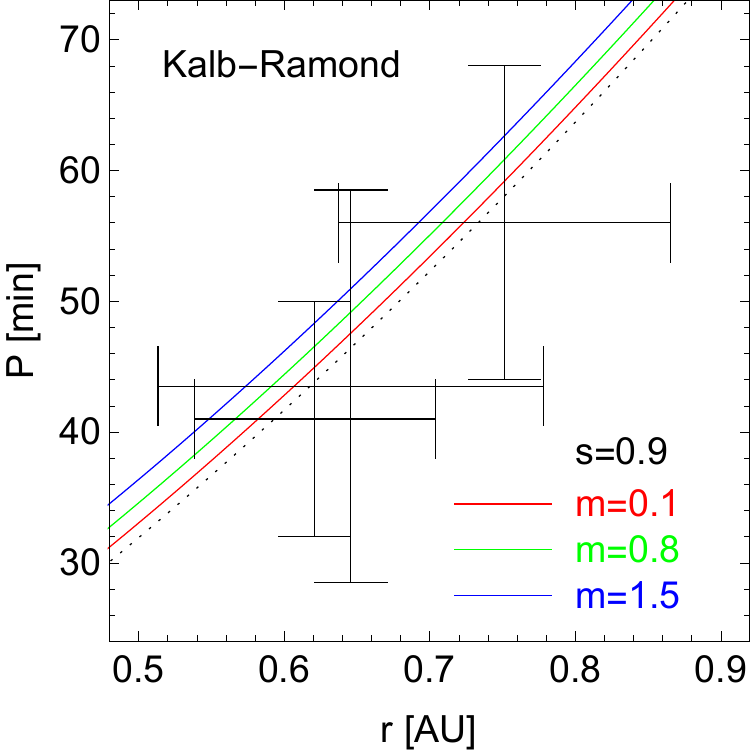}\qquad
\includegraphics[width=0.3\hsize]{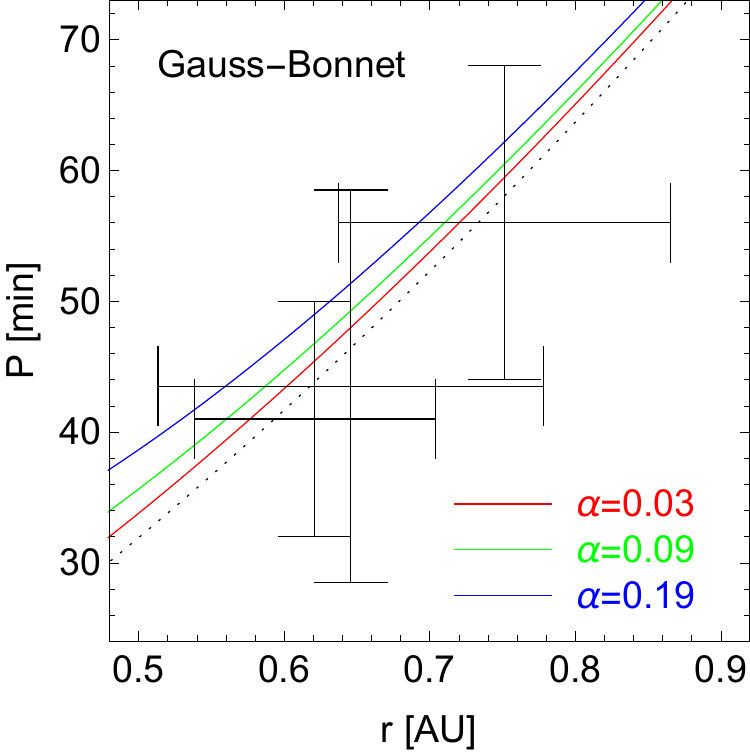}\qquad
\includegraphics[width=0.3\hsize]{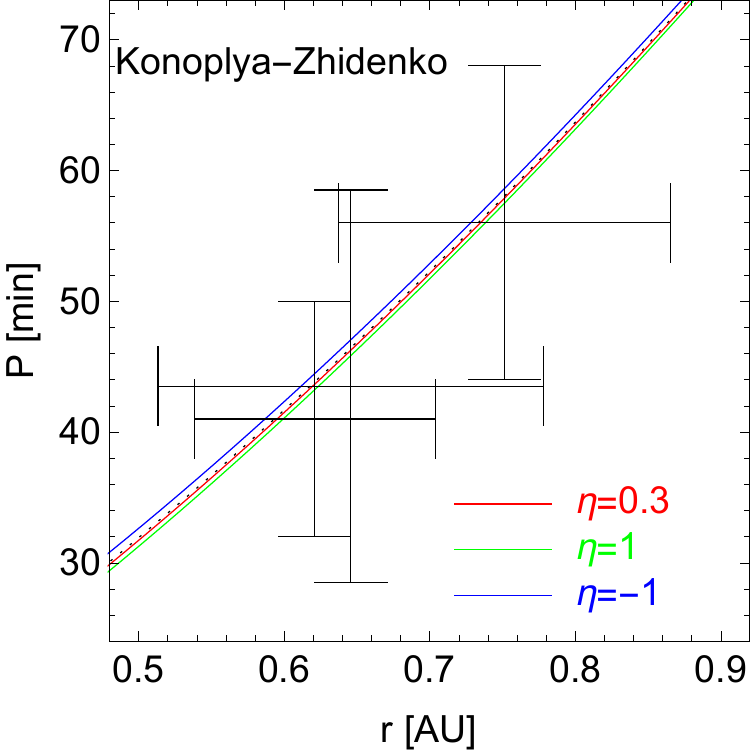}
\includegraphics[width=0.3\hsize]{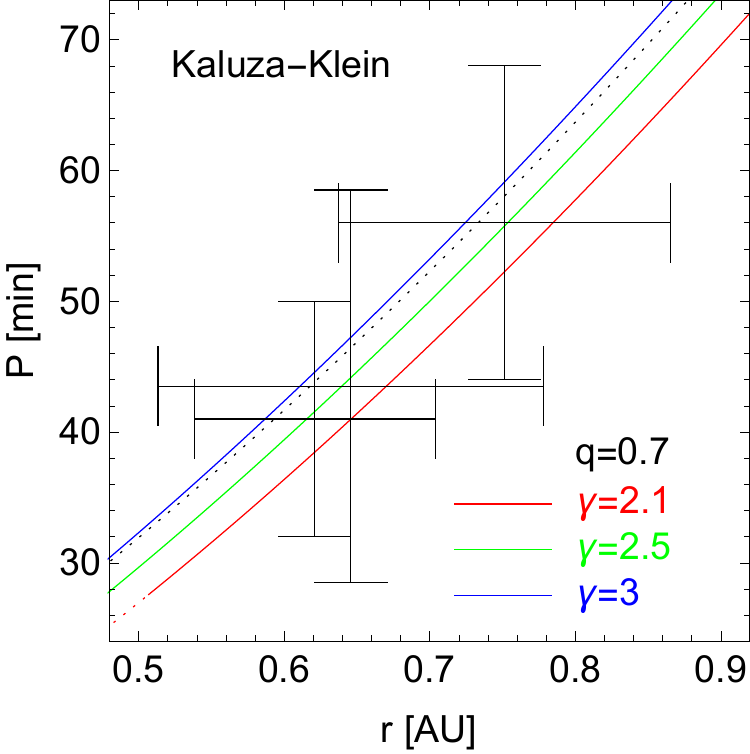}\qquad
\includegraphics[width=0.3\hsize]{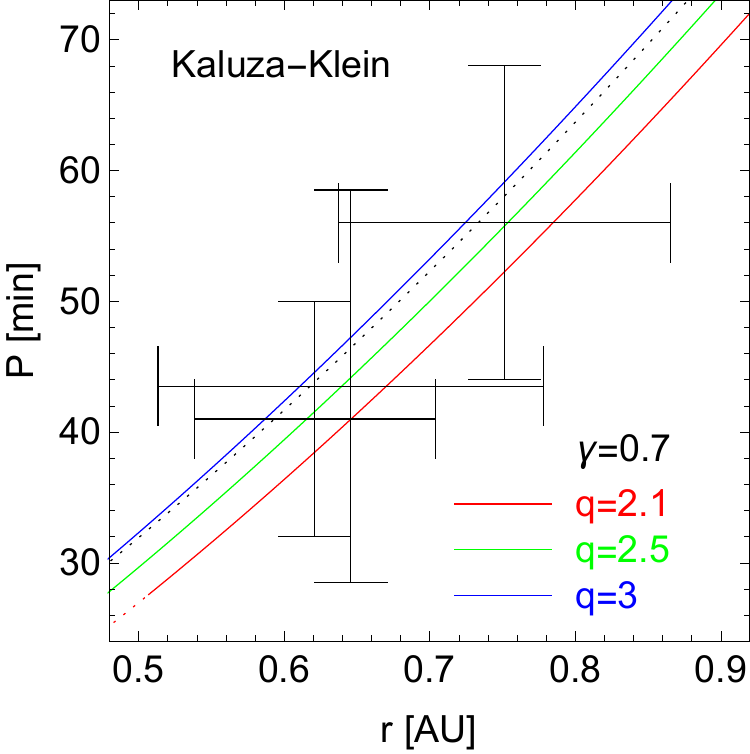}\qquad
\includegraphics[width=0.3\hsize]{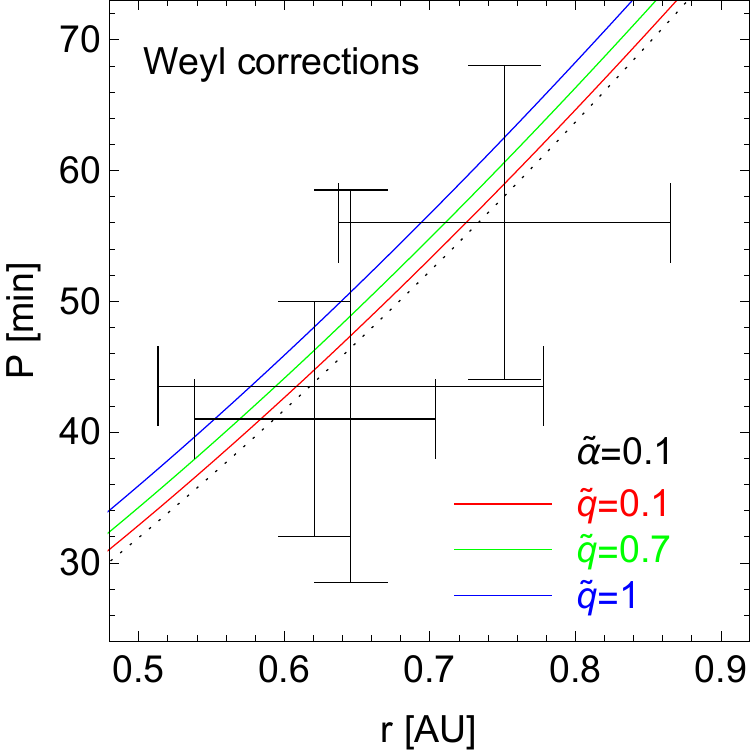}
\caption{continued...}
\end{figure*}
\renewcommand{\thefigure}{2}
\begin{figure*}
\centering
\includegraphics[width=0.3\hsize]{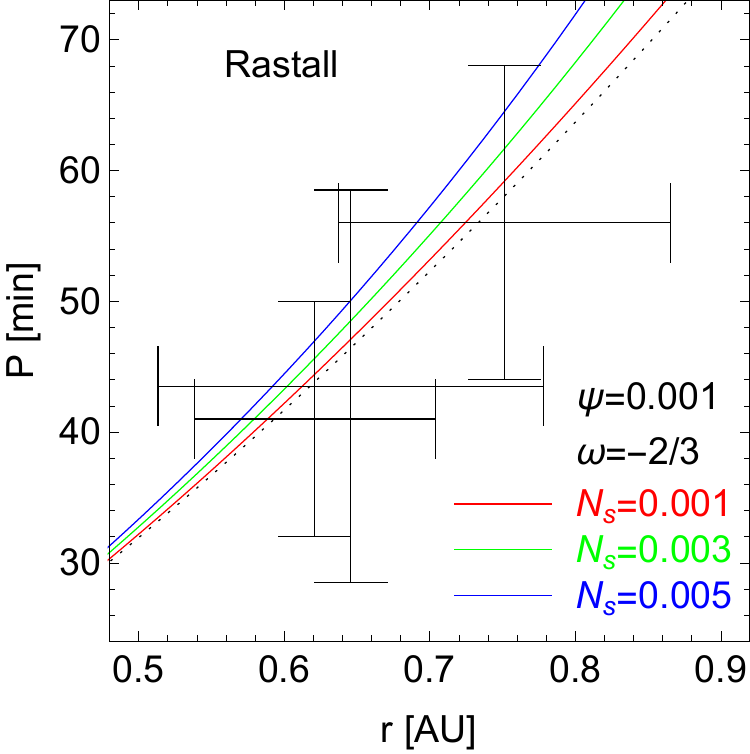}\qquad
\includegraphics[width=0.3\hsize]{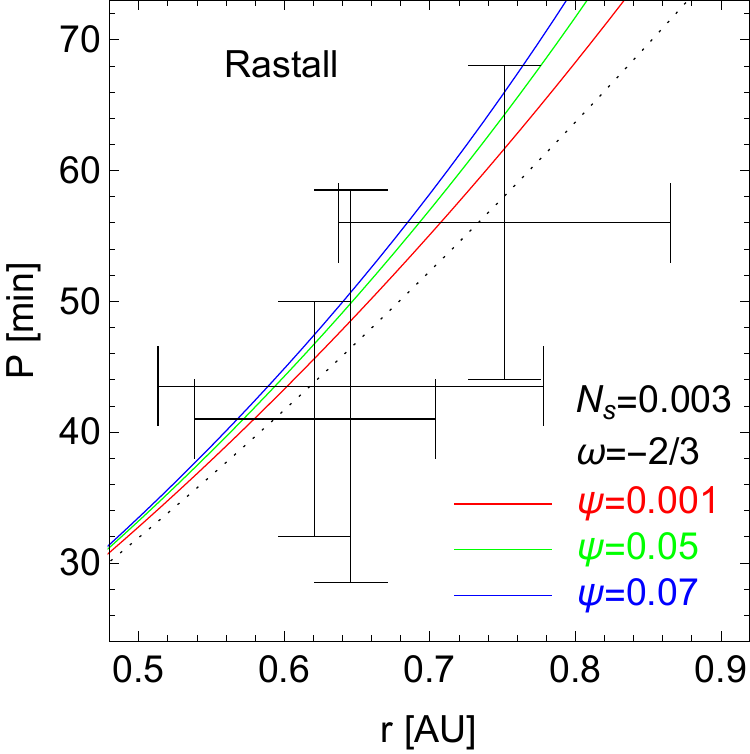}\qquad
\includegraphics[width=0.3\hsize]{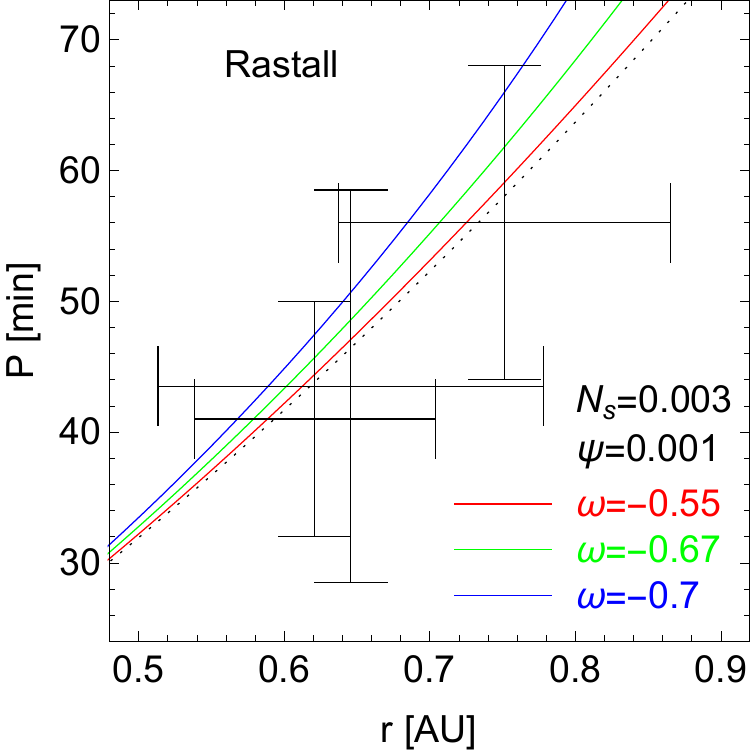}
\includegraphics[width=0.3\hsize]{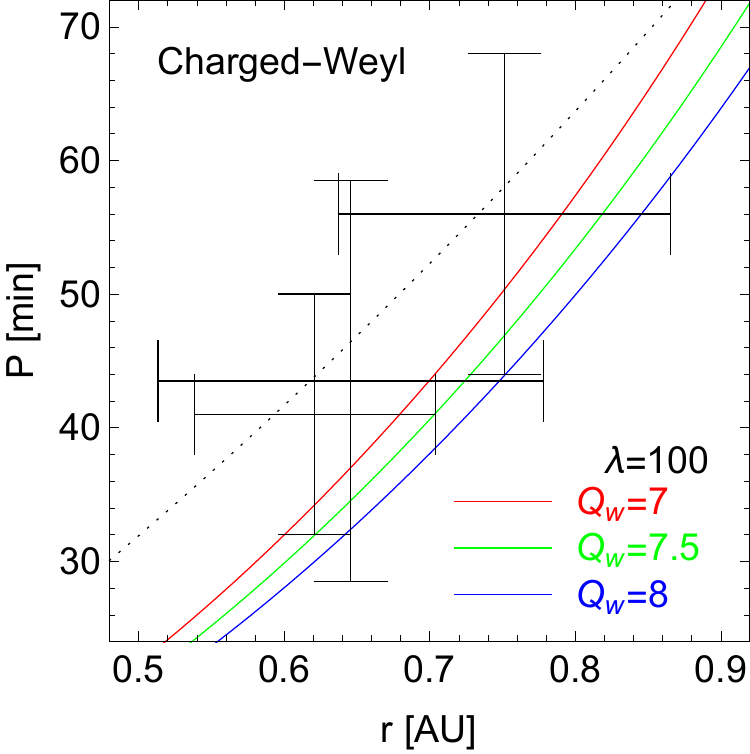}\qquad
\includegraphics[width=0.3\hsize]{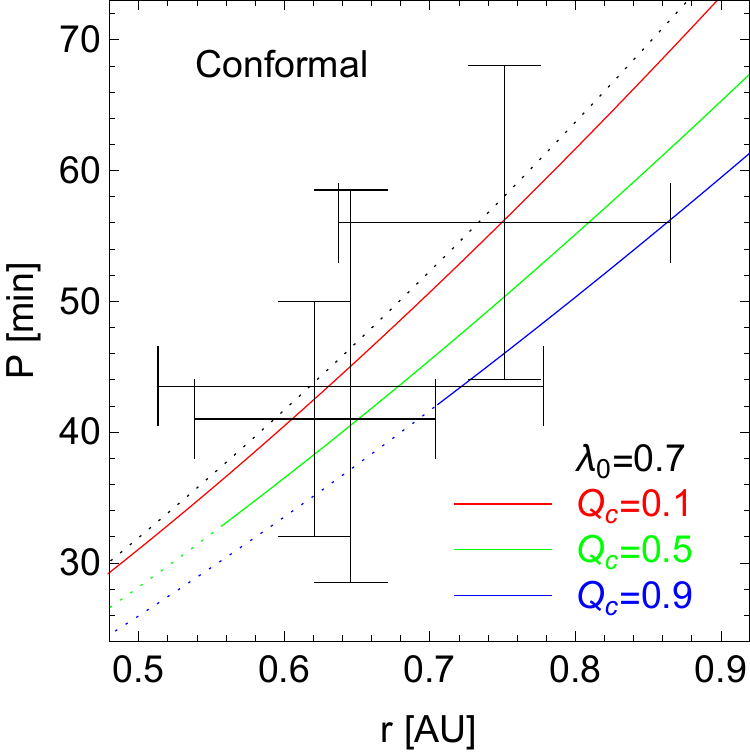}\qquad
\includegraphics[width=0.3\hsize]{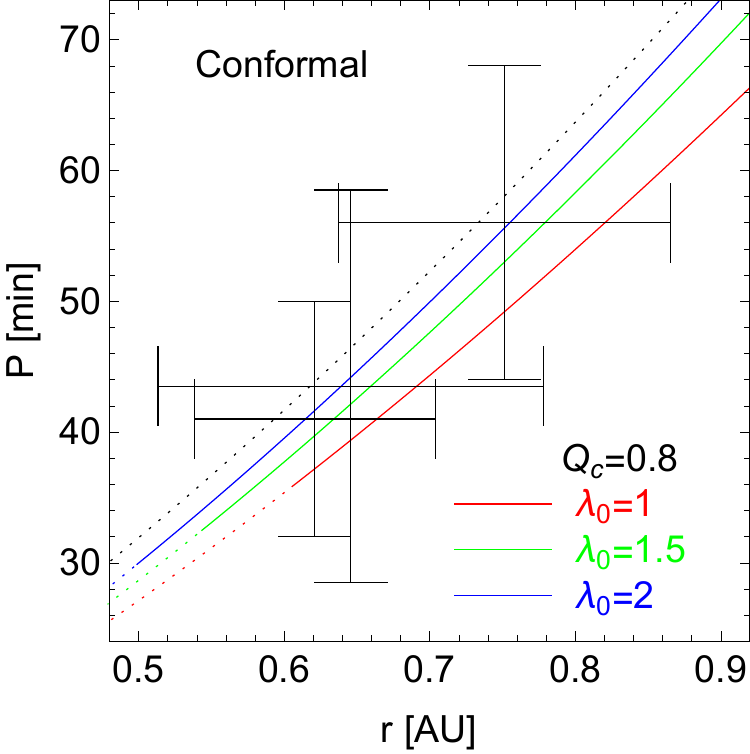}
\includegraphics[width=0.3\hsize]{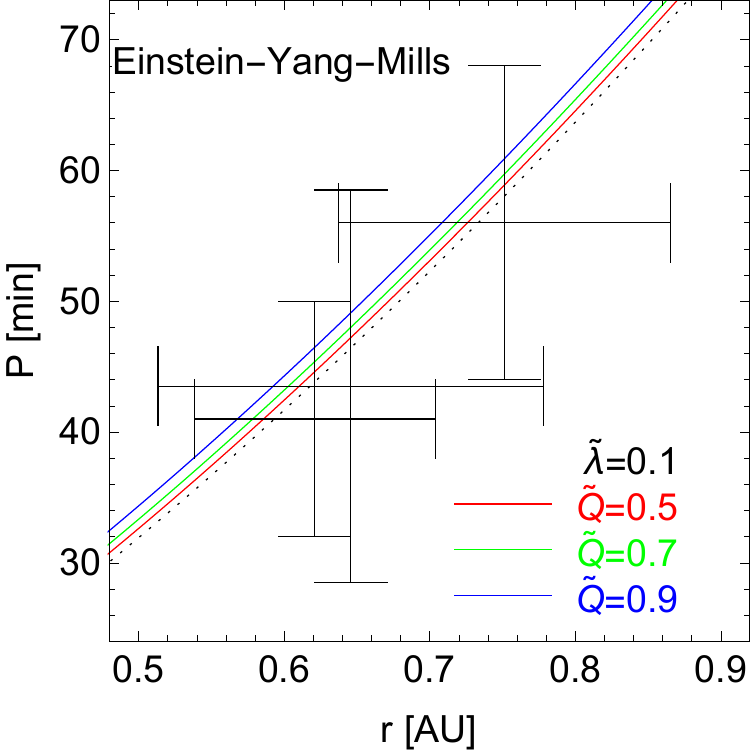}\qquad
\includegraphics[width=0.3\hsize]{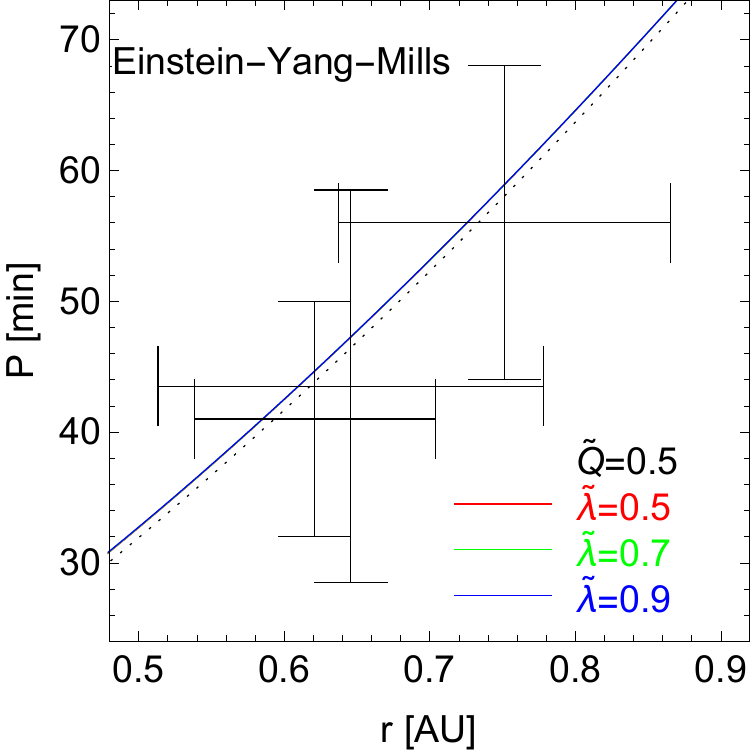}\qquad
\includegraphics[width=0.3\hsize]{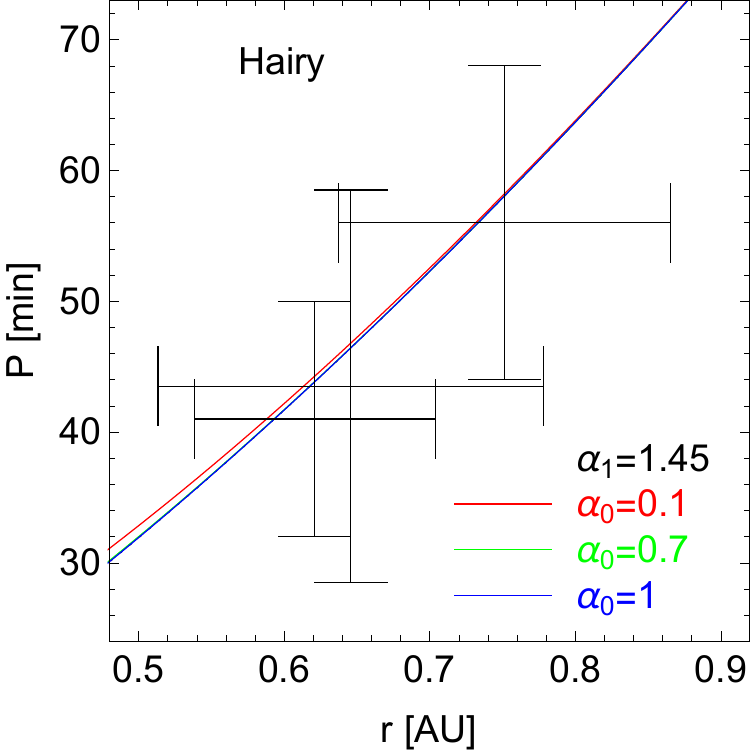}
\includegraphics[width=0.3\hsize]{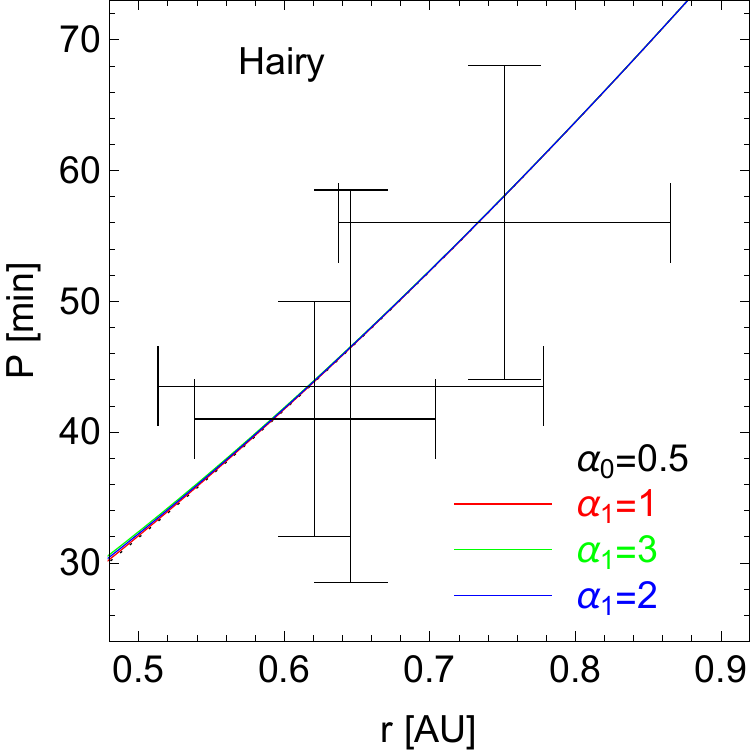}\qquad
\includegraphics[width=0.3\hsize]{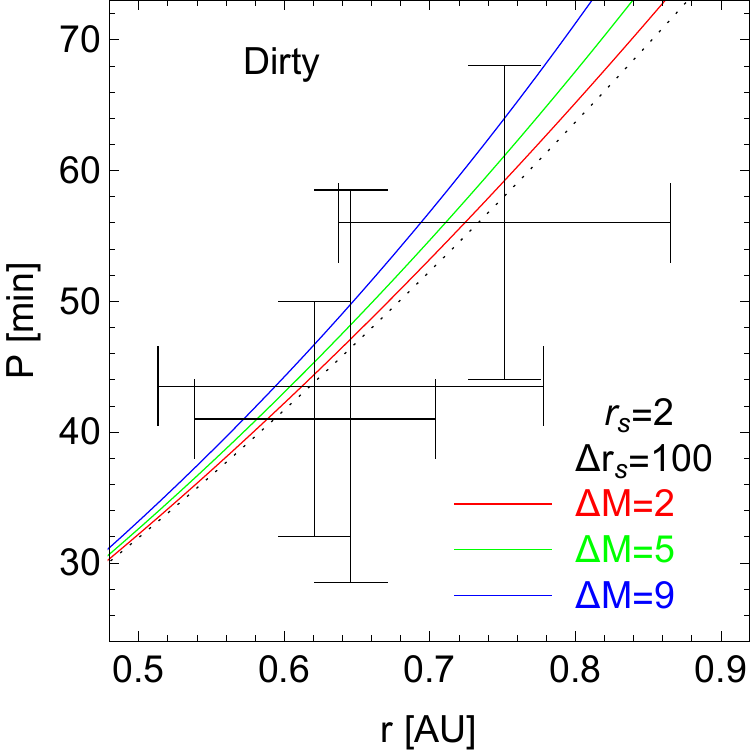}\qquad
\includegraphics[width=0.3\hsize]{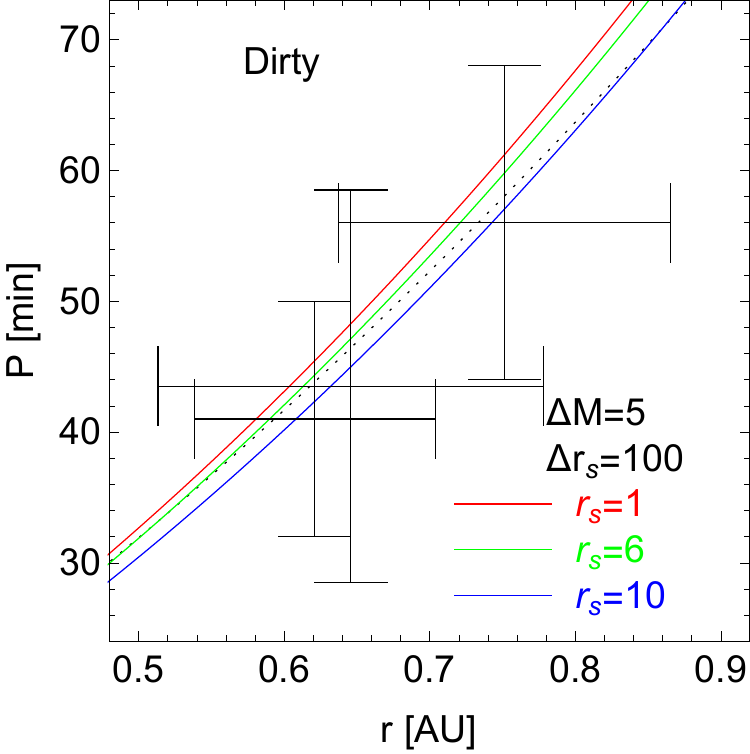}
\caption{continued...}
\end{figure*}
%
\renewcommand{\thefigure}{2}
\begin{figure*}
\centering
\includegraphics[width=0.3\hsize]{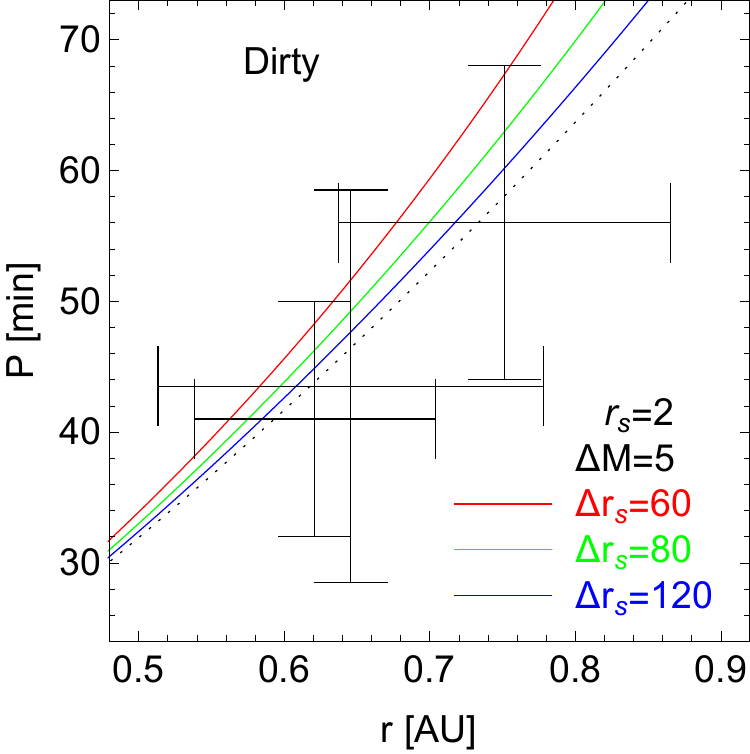}\qquad
\includegraphics[width=0.3\hsize]{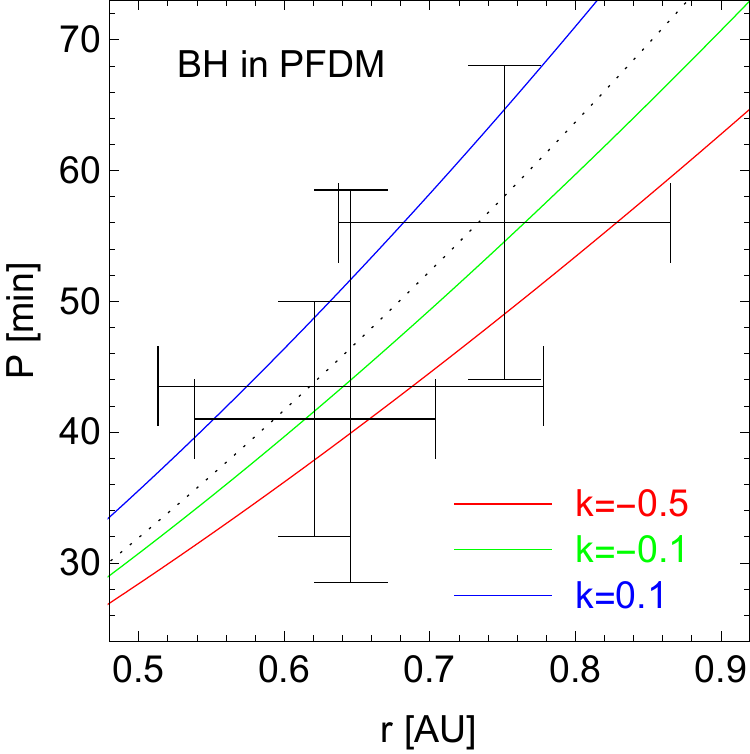}\qquad
\includegraphics[width=0.3\hsize]{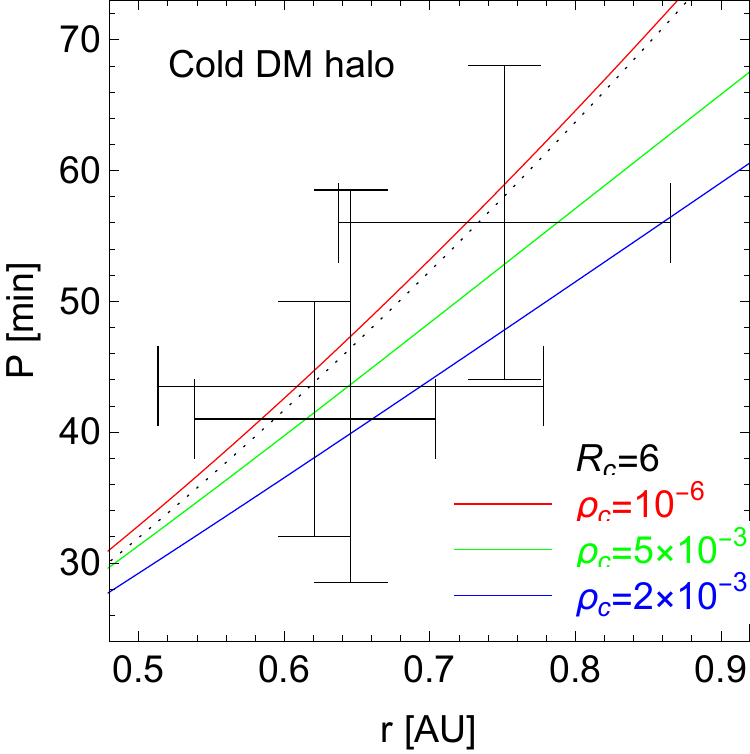}
\includegraphics[width=0.3\hsize]{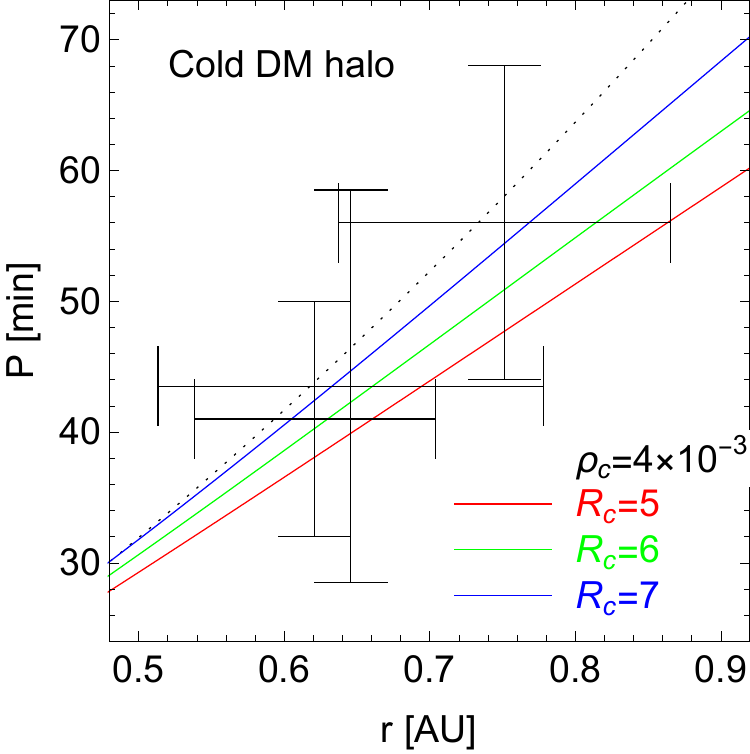}\qquad
\includegraphics[width=0.3\hsize]{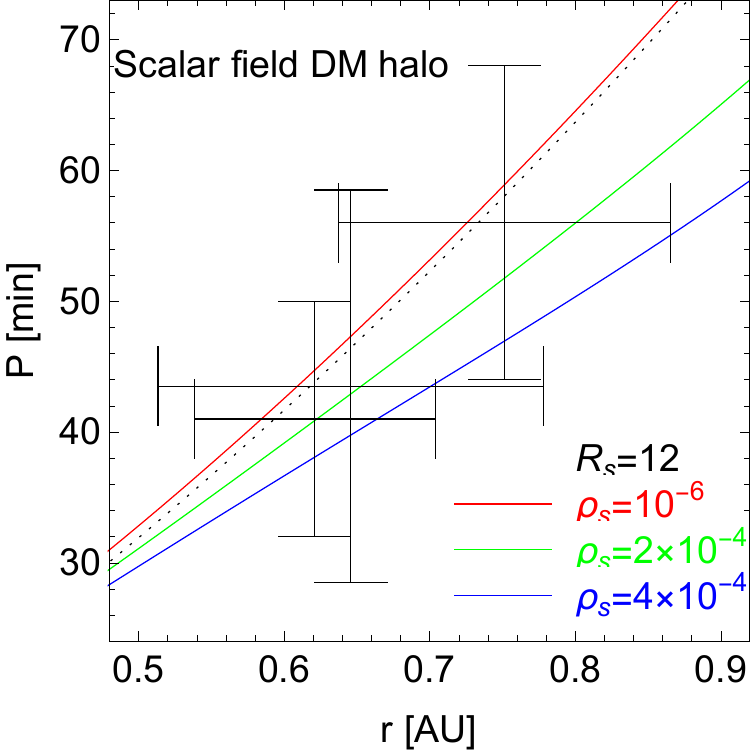}\qquad
\includegraphics[width=0.3\hsize]{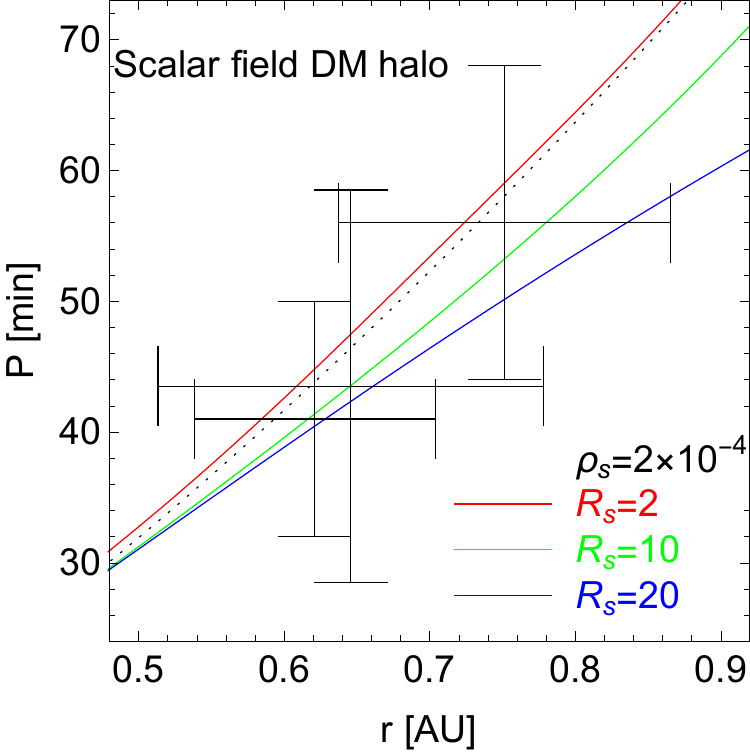}
\includegraphics[width=0.3\hsize]{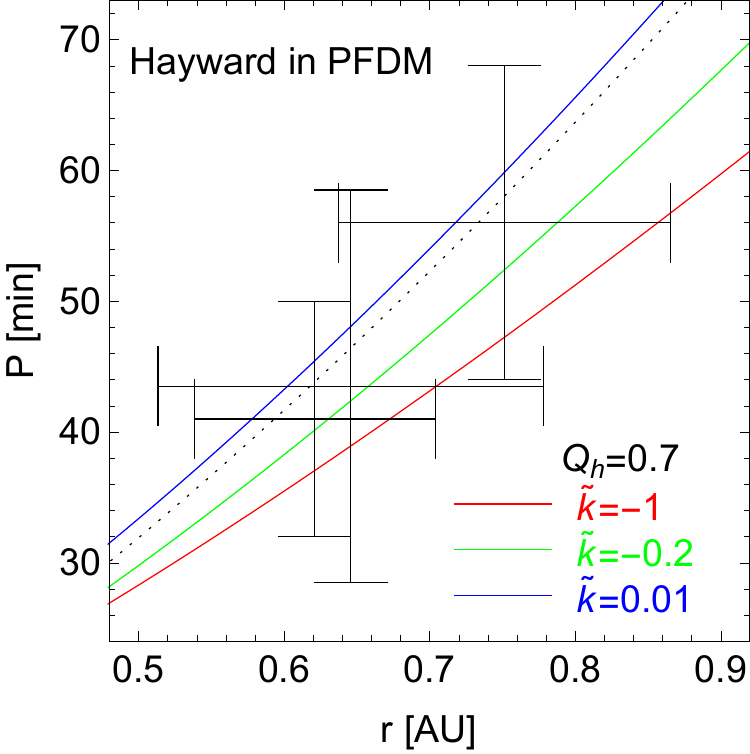}\qquad
\includegraphics[width=0.3\hsize]{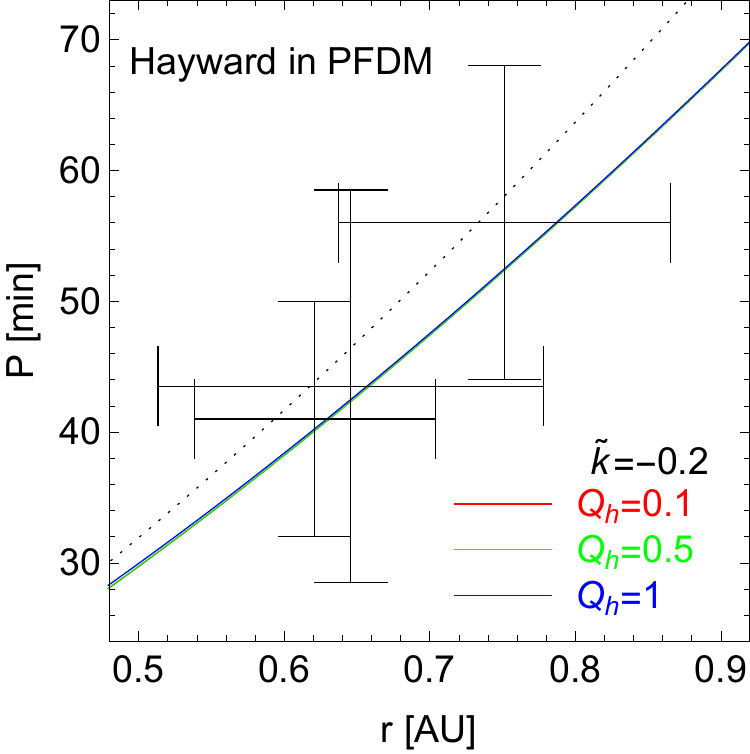}\qquad
\includegraphics[width=0.3\hsize]{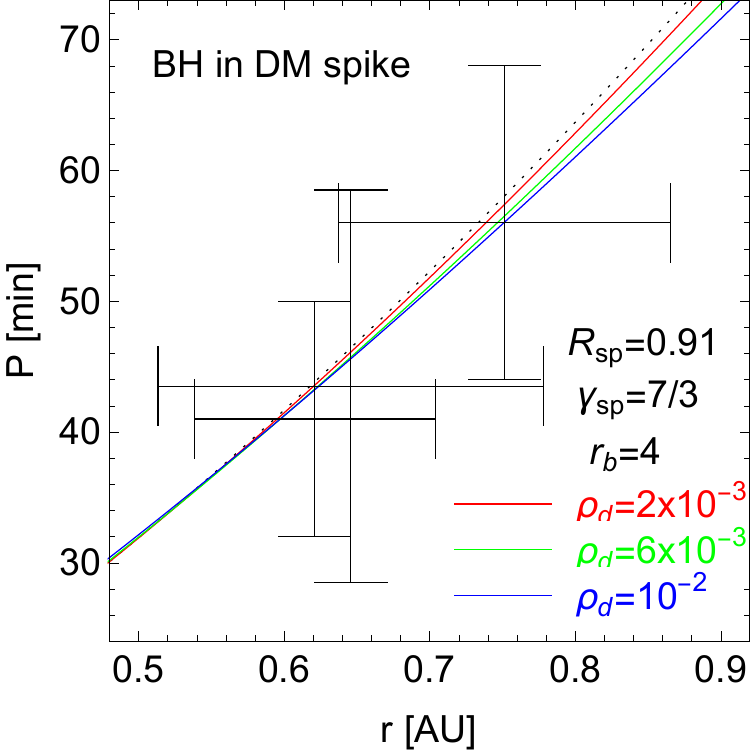}
\includegraphics[width=0.3\hsize]{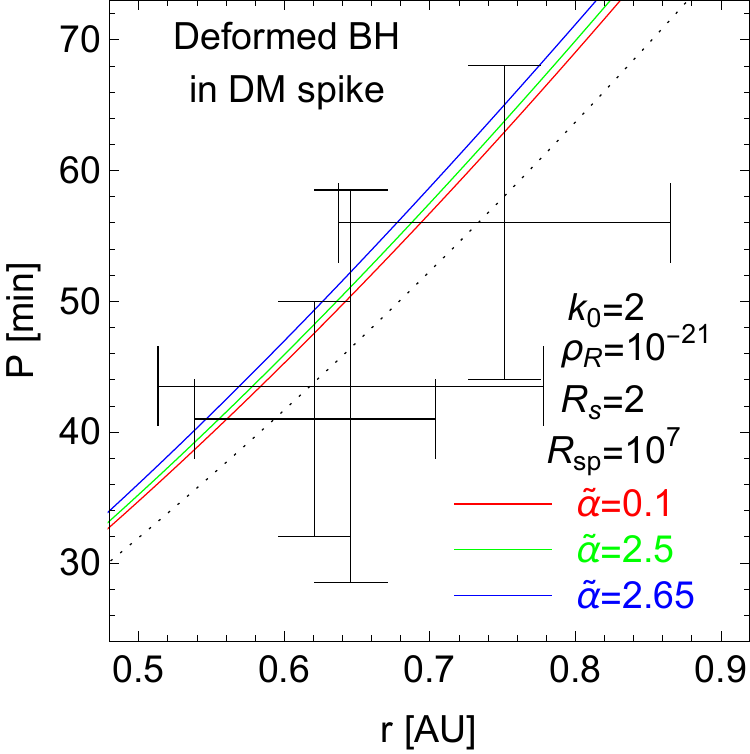}\qquad
\includegraphics[width=0.3\hsize]{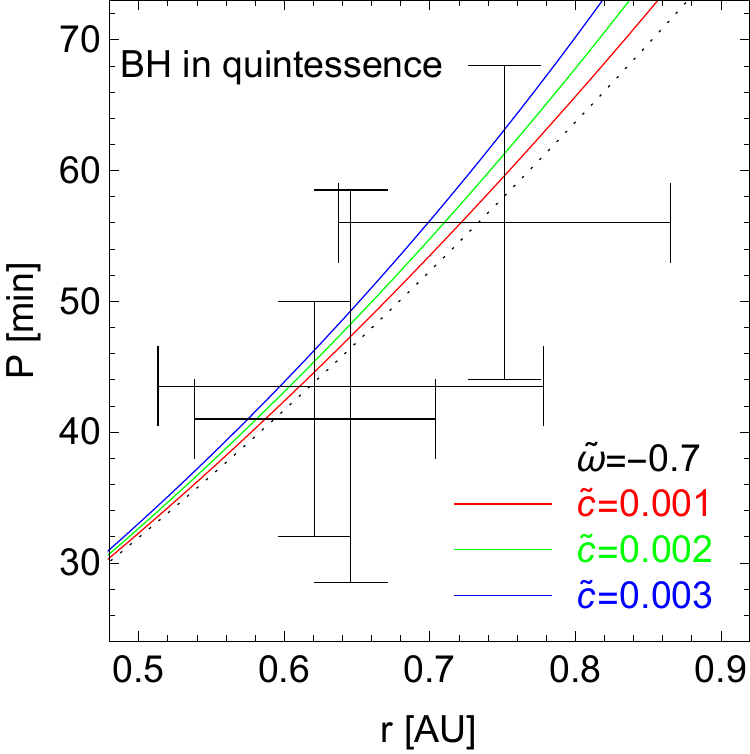}\qquad
\includegraphics[width=0.3\hsize]{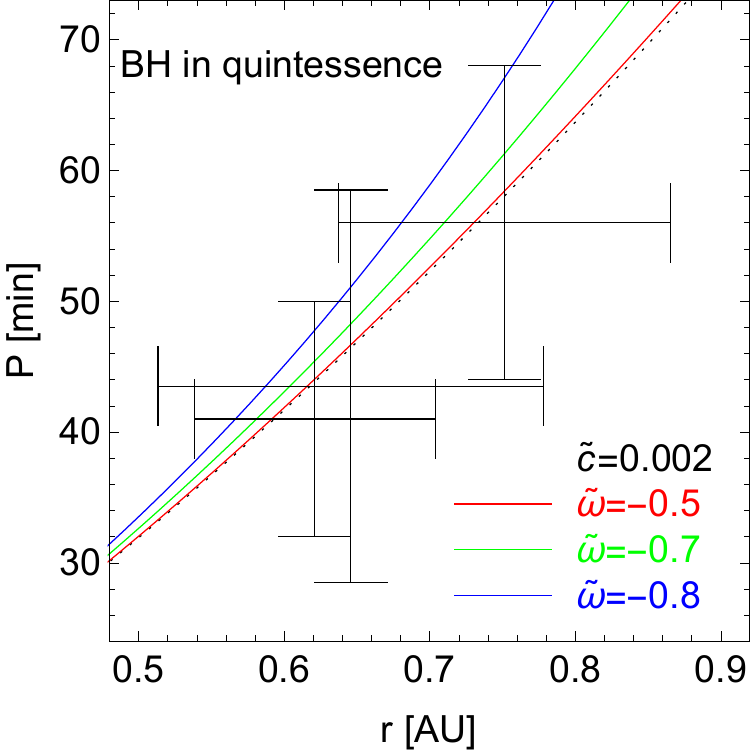}
\caption{continued...}
\end{figure*}

\section{Fitting to the flare data}

Supermassive BH located at the dynamical center of our Galaxy has estimated mass of $M\sim4\times10^6\Msun$ from stellar dynamics around this gravitating center \citep{Do-etal:2013:APJL:}, there is also more recent estimation to $M=4.30\times 10^6\Msun $ with a precision of about $\pm0.25\%$ \citep{GRAVITY:2021:AAA:}. The most prominent S2 star is orbiting Galactic center with pericenter at $r_{\rm p}=120\mskip3mu{\rm AU}$, which is bigger then $10^3$ in geometrical units. The Galactic BH horizon diameter, which in geometrical units should be smaller then $4$, is $0.3\mskip3mu{\rm AU}$ only. Hence the mass $M=4.3\times10^6\Msun$ measured be orbital dynamics of these S2 stars can be viewed as Newtonian far field limit on mass of relativistic compact object located at Galactic center. Hot-spot flares are observed from region much closer to the center, at $r\sim0.7\mskip3mu{\rm AU}$ (Venus orbit); with orbital period less then 60 minutes they are on fully relativistic trajectories, enabling us to explore true nature of our Galactic Center.

From formula (\ref{period}), we notice that hot-spot period linearly depends on BH mass, see period-radius plots at Fig.~\ref{Kerr-Fig}. In the case of standard Kerr metric, all three BH masses $M=(3,4.3,7)\times10^6\Msun$ fit the observed hot-spot flare data well, and the best fit can be seen for mass around $M\sim4.5\times10^6\Msun$ depending on BH spin. For large value of mass parameter, the circular orbits are situated above the center of error bars, while for small value of $M$, the circular orbits shift below the center. If BH mass $M=4.3\times 10^6\Msun$ will be used, we are able to fit observed data quite well, but still the hot-spot period appear to be little bit above center of error bars and one can conclude that the hot-spot is moving little bit faster then particle on geodesic in Kerr BH spacetime.

Both co-rotating as well as counter-rotating circular orbits of neutral hot-spot in the background of Kerr BH with mass $M=4.3\times10^6\Msun$ are depicted in Fig.~\ref{Kerr-Fig}. Orbiting periods around central BH are influenced by BH spin, but not strongly - all orbits fit the observed position and period of the three flares observed by GRAVITY. We compare the co-rotating ($a=1$) as well as counter-rotating ($a=-1$) orbits with the Schwarzschild ($a=0$) case and observe that the co-rotating orbits of hot-spot orbiting Kerr BH are situated above the orbits around Schwarzschild BH while the counter-rotating orbits lie below it. The dotted parts of the curves are plotted for regions below the ISCO position, while the solid part of curves show the behavior above the ISCO. If one would assume the hot-spot flare to be on stable circular orbit only, then counter-rotating orbits are not the option for higher BH spin ($a<-0.4$) since the ISCO is located too far away for counter-rotating orbit. On the other hand, if one assumes the hot-spot could be also on unstable orbit below ISCO radius, then counter-rotating orbits hot-spot orbits around mildly rotating BH $a=-0.4$ are giving better fits then co-rotating orbits.

Within this article, we consider some alternative BH spacetimes deviating from standard Kerr BH on observed radius-period data. We choose the BH mass $M=4.3\times10^6\Msun$, spin parameter $a=0.4$, and for various different BH spacetimes, we fit the observed positions as well as periods of all three flares observed by GRAVITY in Fig.~\ref{fig_all}. Large variety ($ 35 $ in total) of rotating BH spacetimes representing modifications of the standard Kerr geometry due to alternative gravity theories, or due to combinations of the standard GR vacuum Kerr spacetime with additional influences on the spacetime structure, we give the orbital period-radius relations for the BH with assumed fixed mass $M=4.3\times10^6\Msun$ and dimensionless spin $a=0.4$, with some representative values of the additional parameters representing the role of modifications of vacuum Kerr spacetime.

Total number of BH spacetimes used in this article is $35$ so far. Only $31$ plots of fits have been made, so there are $4$ spacetimes, i.e, KN, braneworld BHs, dyonic charged BHs, and Kerr-MOG BHs, with the same orbital frequency as classical Kerr BH, hence they are not included in period-radius diagram (Fig.~\ref{fig_all}). The period-radius plots shown in Fig.~\ref{fig_all} can give us some limitations on these alternative BH spacetime parameters where orbital period will be lowered down when such new spacetime parameters are introduced. We thus can immediately see the tendencies to increase or decrease the data fitting, being able to select the most favourable modified Kerr spacetimes. The Kaluza-Klein BHs, charged Weyl BHs, regular BHs and the BHs modified by DM field demonstrate clearly the best tendency to fit the data corresponding to observed three flares among all the considered BHs.

\section{Conclusions} \label{kecy}

The detection of three bright ``flares" in the neighbourhood of Galactic center supermassive BH that exhibited the orbital motion at a distance of about $6-11$ gravitational radii from a $4.3\times10^6\Msun$ BH has been declared by the near-infrared GRAVITY@ESO observations at $2.2\,\mu{\rm m}$. The Ks-band observations also disclose that the flares are related with the orbiting luminous mass/hot-spots. We explore the dynamics of neutral hot-spot in the background of various stationary, axisymmetric, and asymptotically flat spacetimes, with the help of the three flares observed by GRAVITY on May~27, July~22, and July~28, 2018. We compared the co/counter-rotating equatorial hot-spot circular orbits around classical Kerr BH with different masses.


The equatorial circular orbits of hot-spot for most of the considering BHs are situated above the center of error bars of all three flares if we assume central object with mass $M=4.3\times10^6\Msun$ and dimensionless spin $a=0.4$, see Fig.~\ref{fig_all}. To obtain better fit of observed data, one can increase the central BH mass to $M=5\times10^6\Msun$ or one can use some non-Kerr BH spacetime. In such alternative BH spacetimes new free parameters are introduced, which could be used to obtain better fit to observed GRAVITY data. The period-radius diagrams in Fig.~\ref{fig_all} can give us some limitations on these alternative BH spacetime parameters where orbital period will be lowered down when such new spacetime parameters are introduced. Our results indicate that as favourable candidates can be considered the Kaluza-Klein BHs in Kaluza-Klein theory, charged Weyl BHs in Weyl gravity, regular BHs in conformal massive gravity and the BHs modified in DM field that are demonstrating the best tendency to fit the data corresponding to observed three flares among all the considered BHs.

All equatorial circular orbit period-radius relations of hot-spot orbiting the BHs, considering here both GR as well as modified theories of gravity, fit the observed position and period of all three flares observed by GRAVITY and it is hard to give conclusive answer on which alternative BH spacetime is correct - obviously more observations will be needed. However, it is instructive that between the best candidates belong all the considered cases of BHs influenced by DM in their vicinity. We also would like to note that the possible role of the electromagnetic interaction of slightly-charged hot spot (or slender torus) with magnetized Kerr BH has been considered in \citep{Tur-etal:2020:APJ:,Kar-etal:2021:COSPAR:}. Such non geodesic effects can also have strong influence on hot-spot dynamics and they are not considered in the article.


\section*{Acknowledgments}

The authors would like to thank Michal Zaja\v{c}ek for very useful discussion regarding the astrophysical nature of our Galactic Center. The authors M.K. and Z.S. would like to express their acknowledgments for the Research Centre for Theoretical Physics and Astrophysics, Institute of Physics, Silesian University in Opava, Z.S. acknowledge the Czech Science Foundation Grant No. 19-03950S.

\bibliography{reference}{}
\bibliographystyle{aasjournal}

\end{document}